\documentclass[traditabstract]{aa} 
\usepackage{txfonts}
\usepackage{supertabular}
\usepackage{graphics,graphicx}

\begin{document}

\title{Size growth of red-sequence early-type galaxies in clusters in the last
10 Gyr\thanks{Table 3 is only available in electronic form
at the CDS via anonymous ftp to cdsarc.u-strasbg.fr (130.79.128.5)
or via http://cdsweb.u-strasbg.fr/cgi-bin/qcat?J/A+A/}.}
\titlerunning{The mass-size relation up to $z=1.8$} 
\author{S. Andreon\inst{1} \and 
Hui Dong\inst{2} 
\and A. Raichoor\inst{3} 
}
\authorrunning{Andreon et al.}
\institute{
$^1$ INAF--Osservatorio Astronomico di Brera, via Brera 28, 20121, Milano, Italy,
\email{stefano.andreon@brera.inaf.it} \\
$^2$ Instituto de Astrofisica de Andalucia (CSIC), Glorieta de la Astronoma S/N, E-18008 Granada, Spain \\
$^3$ CEA, Centre de Saclay, IRFU/SPP, 91191 Gif-sur-Yvette, France
\\ 
}
\date{Accepted ... Received ...}
\abstract{
We carried out a photometric and structural analysis in the rest-frame
$V$ band of a mass-selected ($\log M/M_\odot >10.7$) sample of 
red-sequence galaxies in 14 galaxy clusters, 6 of which  are at
$z>1.45$, namely JKCS041, IDCS J1426.5+3508, SpARCS104922.6+564032.5,
SpARCSJ022426-032330, XDCPJ0044.0-2033, and SPT-CLJ2040-4451. To this
end, we reduced/analyzed about 300 orbits of  multicolor images taken 
with the Advanced Camera for Survey and the Wide Field Camera 3 on the
Hubble Space Telescope.  We uniformly morphologically classified
galaxies from $z=0.023$ to $z=1.803$, and we homogeneously derived
sizes (effective radii) for the  entire sample. Furthermore, our size
derivation allows, and therefore is not biased by,  the presence of
the usual variety of morphological structures seen in early-type
galaxies, such as bulges, bars, disks, isophote twists, and ellipiticy
gradients. By using such a mass-selected  sample, composed of 244
red-sequence early-type galaxies, we find that the $\log$ of the
galaxy size at a fixed stellar mass, $\log M/M_\odot= 11$ has increased
with time at a rate of $0.023\pm0.002$ dex per Gyr over
the last 10 Gyr, in marked contrast with the threefold increase found
in the literature for galaxies in the general field over the same period.
This suggests, at face value, that secular processes should be excluded
as the
primary drivers of size evolution because we observed an environmental
environmental dependent size growth. Using spectroscopic ages of Coma early-type galaxies we
also find that recently quenched early-type galaxies are a numerically
minor population not different enough in size to alter the mean size
at a given mass, which implies that the progenitor bias is minor, i.e.,
that the size evolution measured by selecting galaxies at the redshift of
observation is indistinguishable from the one that compares ancestors
and descendents.
}

\maketitle

\section{Introduction}

In $\Lambda$CDM-based models of galaxy formation, massive
galaxies are expected to assemble hierarchically from mergers of smaller
systems. Such models predict that the progenitors of today's most massive galaxies
formed stars in situ at $z\gtrsim 2-3$ in intense, short-lived bursts, while
most of the subsequent mass assembly occurred via the accretion of stars formed in
smaller systems (e.g., De Lucia \& Blaizot 2007; Khochfar \& Silk 2006). 
These predictions are supported
by observations of the slowly evolving mass-to-light ratios of massive
early-type galaxies (Treu et al. 2005) and by their high $[\alpha/Fe]$ 
abundances (Thomas et al. 2005). 

A striking result that demands explanation, however, is that quiescent massive galaxies 
at $z = 2.5$ seem typically $3-5\times$ smaller in physical size than local galaxies of 
equal stellar mass (see Newman et al. 2012 and references therein). Remarkably, about 40\% 
of the expansion occurs within the redshift interval $z = 1.5 - 2.5$, corresponding to 
a period of only 1.6 Gyr. Apparently, this represents a key evolutionary period when 
the progenitors of today's massive elliptical galaxies matured rapidly. 

The most commonly posited mechanism for this growth is merging. Although
major mergers involving near-equal mass galaxies are relatively rare and
inefficient, minor mergers involving low-mass companions provide a promising
explanation. Both simulations (e.g., Naab et al. 2009) and observations (Hopkins et 
al. 2009, Bezanson et al. 2009) suggest that stellar material accreted in minor mergers 
remains largely at large radii and contributes more to the growth in size than to the stellar 
mass (van Dokkum et al. 2010). Although the observed rate of minor mergers measured 
in deep CANDELS data appears sufficient to explain the size growth over $0<z<1.5$
(Newman et al. 2012), it seems unable to account for the rapid growth during the
formative period around $z\simeq2$. Thus, additional processes may be
required to complete our understanding of the assembly history of massive
quiescent galaxies since $z>2$.

Environmental trends offer a powerful way to make progress. In a
merger-driven scenario we expect a strong dependence between the assembly
rate and the local environment because galaxies inhabiting denser regions
should experience more frequent mergers and demonstrate a faster growth rate
(e.g., McIntosh et al. 2008). On the other hand, if secular processes are
primarily responsible -- for example, in models where adiabatic expansion
results from mass expelled in stellar winds or by AGN (e.g., Fan et al. 2008,
Damjanov et al. 2009, Ishibashi et al 2013) -- then the assembly rate should be 
mostly independent of environment. Much of the enhanced merging activities in 
dense systems should occur during the proto-cluster phase at high redshift 
when infalling groups have not yet virialized  (e.g., Lotz et al. 2011). Evolved 
clusters have presumably progressed beyond the dominant merging phase 
leaving a clear imprint in the structure of the cluster galaxies.

\begin{table*}
\caption{Sample, number of galaxies, and imaging exposure times for clusters at $z>1$}
\begin{tabular}{l r r r r r r r r r r}
\hline
ID & z & $n_{gal}$ & $n_{gal,bkg}$ & \multispan{7}{\hfill $t_{exp}$ [ks] \hfill} \\
\cline{5-11}  
& & & & \multispan{2}{\hfill ACS \hfill} & & \multispan{4}{\hfill WFC3 \hfill} \\
\cline{5-6}  \cline{8-11}  
& & & & F606W& F814W &	& F814W  & F105W &  F140W &  F160W \\
\hline
JKCS041 	        & 1.80 & 7  &  0  & -	& -	& &   -    &   2.6  &	-	 &  {\bf 4.5} \\	 
IDCS J1426.5+3508	& 1.75 & 5  & 1.3   & 21.8  & 12.6  & &   -    &   4.9  &	{\bf 5.4}  &   7.6 \\ 	 
SpARCS104922.6+564032.5 & 1.71 & 3  & 1.3   & -	& -	& &   1.1  &   5.0  &	-	 &  {\bf 5.2} \\	 
SpARCSJ022426-032330   & 1.63 & 3  & 1.1   & -	& -	& &   1.3  &   1.8  &	{\bf 2.0}  &   2.6 \\ 	 
XDCP J0044.0-2033	& 1.58 & 5  & 0.7   & -	& -	& &   1.6  &   4.7  &	{\bf 5.2}  &   2.6 \\ 	 
SPT-CL J2040-4451	& 1.48 & 9  & 0.4   & 2.1	& -	& &   3.3  &   7.2  &	{\bf 8.0}  &   -	 \\	 
ISCS J1432.3+3253       & 1.40 & 3  & 0.4   &  -     &    -   & &   3.7  &    9.3  &  {\bf 8.8}   &    -     \\ 
SPT-CL J0205-5829	& 1.32 & 14  & 0.4   & 7.7	& -	& &  8.2  &   15.0  &	{\bf 14.2}  &   3.3	 \\	 %
\hline
HUDF			&- & 2  & - & 24.3  & 7.0  & & - &  5.5 & {\bf 5.5}  &  5.6 \\
Parallel MACSJ0414	&- & 2  & - & 25.0  & 19.8 & & -   &    17.8 &  {\bf 14.6}  &  14.6 \\    
Parallel MACSJ1149 	&- & 1  & - & 14.6  &  15.1      & & -  &    14.3 &  {\bf 16.1}  &  16.7 \\   
\hline		   
\end{tabular}      				  
\hfill \break
Numbers in boldface identify the band used for the isophotal analysis. 	   
\hfill \break	   
\end{table*}

\begin{table*}
\caption{Sample, number of galaxies, and imaging exposure times for clusters at $z<1$}
\begin{tabular}{l r r r r r r r r r r r l}
\hline
ID & z & $n_{gal}$ & $n_{gal,z-memb}$ & \multispan{8}{\hfill ACS $t_{exp}$ [ks] \hfill} \\
\cline{5-12}
& & & & F435W & F475W & F555W &	F606W & F625W  & F775W &  F814W &  F850LP & \\
\hline
RXJ0152.7-1357     & 0.84 & 21  &  13  &  & &  & & 19.0   & 19.2  &  & {\bf 19.0}  \\      
MACSJ1149.5+2223$^a$   & 0.54 & 22  & 17  & &  & 4.5 & &    &  & {\bf 4.6}/104.2 &    \\  	   
MACSJ1206.2-0847$^a$   & 0.44 & 22  & 16 &  & &   & 6.6 & &    & {\bf 8.5} &     \\      
Abell 2744 	   & 0.31 & 23  & 21 & 45.7 & &  & 23.6  &   &  & {\bf 104.3} &   \\  	  
Abell 2218	   & 0.17 & 21 & 20 & 7.0 & 5.6 & 7.0 & & {\bf 8.4} & 10.7 \\
Abell 1656$^b$ (Coma) & 0.02 & 86 & 86	    &   &  & &    &  &  &     \\   
\hline		   
\end{tabular}      				  
\hfill \break
$^a$ 16-band photometry from CLASH \\
$^b$ photometry, morphology, and effective radii from Andreon et al. (1996,1997) \\
Numbers in boldface identify the band used for the isophotal analysis.	   
\hfill \break	   
\end{table*}

Recognizing the advantage of contrasting size growth
as a function of environment in the critical redshift interval $1.5<z<2.5$,
Hubble Space Telescope (HST) data on the $z=1.803$ cluster JKCS\,041 (Newman et al. 2014)
revealed similar mass-age relations for an unprecedented large 
sample of 15 spectroscopically confirmed quiescent cluster galaxies 
when compared with a similarly selected field sample (Whitaker et al. 2013). 
While the quenching of star formation to produce quiescent
systems is certainly more widespread in early clusters, it apparently proceeds 
at a pace independent of the environment (Newman et al. 2014). Regarding the question of 
environmentally dependent size growth to local galaxies, an intriguing difference 
was found: an upper limit of $0.22$ dex growth  was found from JKCS\,041 to the Coma cluster 
in comparison with a measured variation of 
$0.47$ dex for field galaxies over the same interval (Andreon et al. 2014). This supports 
accelerated growth in dense environments, such as would be the case for merging;
however, a large sample is called for, in particular to confirm the effect 
currently based on just two clusters.

Quiescent galaxies at both high and low redshift have various
structural components:
high-redshift early-type galaxies are known to have a disk (e.g., van der Wel et al.
2011; Chang et al. 2013; Lang et al. 2014), i.e., not to be single-component galaxies,
whereas local early-type galaxies are known to have bars, disks,
ellipticity, and position angle gradients (e.g., Jedrzejewski et al. 1987, 
Bender \& Moellenhoff 1987; Nieto \& Bender 1989). Therefore, in order
to derive reliable sizes, the model to be fitted to the galaxy surface
brightness should not be an oversimplified
description of nature complexity, for
example one that adopts a radial-independent value for the galaxy
ellipticity and position angle.

In this work, we uniformly analyze a sample of 14 galaxy clusters, 6 of which at $z>1.47$.
The remaining clusters
uniformly sample the last 7 Gyr of the Universe age.
All clusters are observed in the rest-frame $V$ band  and
all galaxies are uniformly morphologically classified. 
Sizes are homogeneously derived for all galaxies and in a way that is unbiased by the
presence of common components of early-type galaxies, such as bulges, disks, and bars.  
The resulting, mass-selected sample of 244 early-type elliptical and lenticular
galaxies allows us to derive the galaxy size
at a fixed stellar mass and how it depends on look-back time.

Throughout this paper, we assume $\Omega_M=0.3$, $\Omega_\Lambda=0.7$, 
and $H_0=70$ km s$^{-1}$ Mpc$^{-1}$. Magnitudes are in the AB system.
We use the 2003 version of Bruzual \& Charlot (2003) stellar population synthesis
models with solar metallicity and a Salpeter initial mass function (IMF).
We use stellar masses that count only the
mass in stars and their remnants.
For a single stellar population, or $\tau=0.1$ Gyr model, the 
evolution of the stellar mass between 2 and 13 Gyr age is about 5\% percent. Therefore,
comparisons (e.g., of radii) at a fixed present-day mass is degenerate with comparisons
with mass at the time of the observations (see Andreon et al. 2006 for a 
different situation).

\begin{figure*}
\centerline{\includegraphics[width=18truecm]{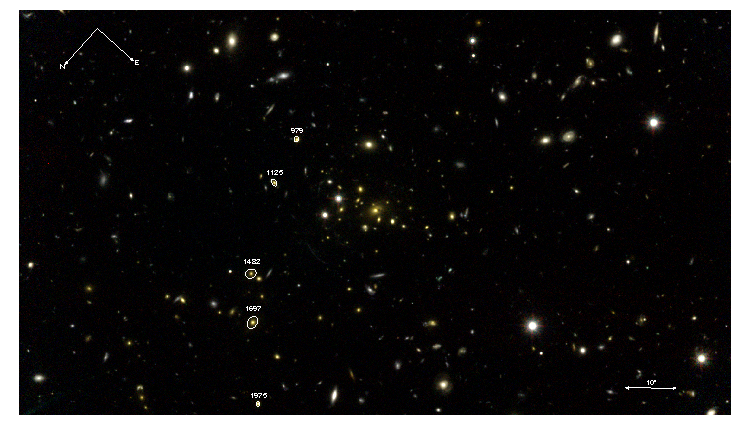}}
\caption[h]{Three-color (F105W-F140W-F160W) image of the $z=1.75$ IDCS J1426.5+3508 cluster. 
Red-sequence, early-type galaxies are marked. The late-type morphology of most
galaxies cannot be appreciated in this figure and requires inspecting the full-resolution
fits image.
}
\end{figure*}

\begin{figure*}
\centerline{\includegraphics[width=18truecm]{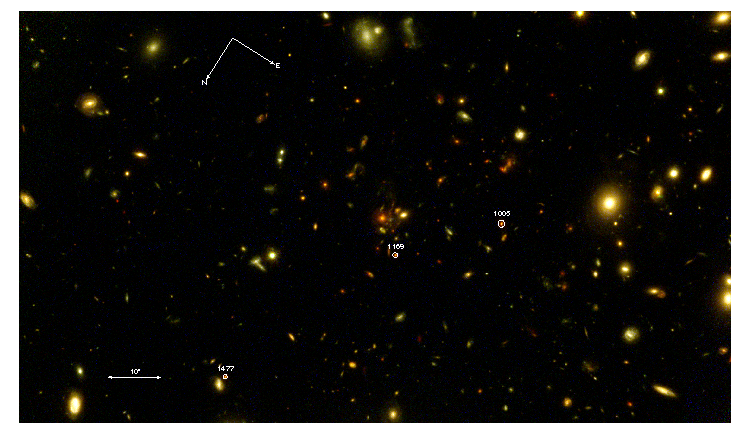}}
\caption[h]{Three color (F814W-F105W-F160W) image of the $z=1.71$ SpARCS104922.6+564032.5 cluster. 
Red-sequence, early-type galaxies are marked. The late-type morphology of most
galaxies cannot be appreciated in this figure.
}
\end{figure*}

\begin{figure*}
\centerline{\includegraphics[width=18truecm]{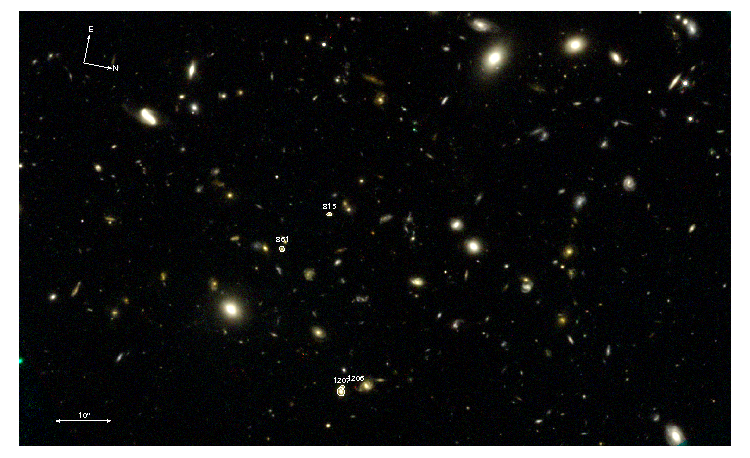}}
\caption[h]{Three color (F105W-F140W-F160W) image of the $z=1.63$ SpARCSJ022426-032330 cluster. 
Red-sequence, early-type galaxies are marked. The late-type morphology of most
galaxies cannot be appreciated in this figure.
}
\end{figure*}

\begin{figure*}
\centerline{\includegraphics[width=18truecm]{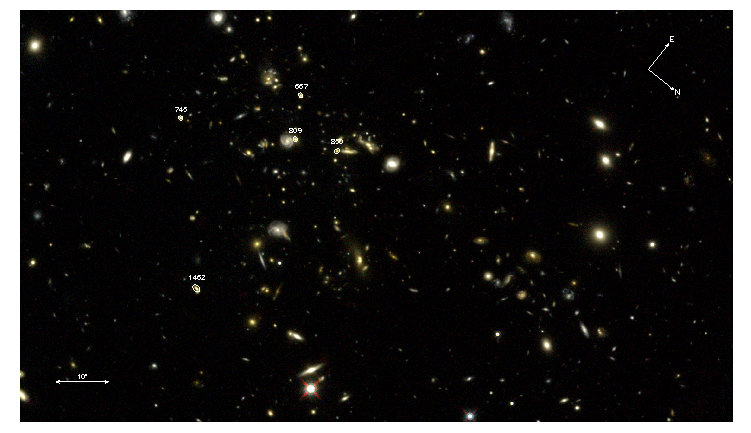}}
\caption[h]{Three color (F105W-F140W-F160W) image of the $z=1.58$ XDCP J0044.0-2033 cluster.
Red-sequence, early-type galaxies are marked. The late-type morphology of most
galaxies cannot be appreciated in this figure.
}
\end{figure*}

\begin{figure*}
\centerline{\includegraphics[width=18truecm]{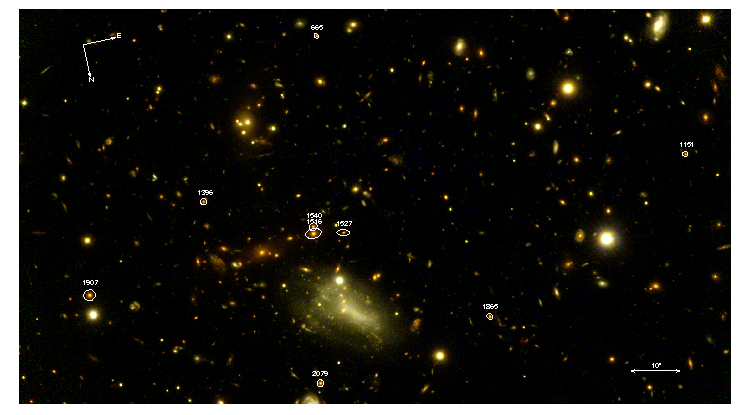}}
\caption[h]{Three color (F814W-F105W-F140W) image of the $z=1.48$ SPT-CL J2040-4451 cluster.
Red-sequence, early-type galaxies are marked. The late-type morphology of most
galaxies cannot be appreciated in this figure.
}
\end{figure*}

\begin{figure*}
\centerline{\includegraphics[width=18truecm]{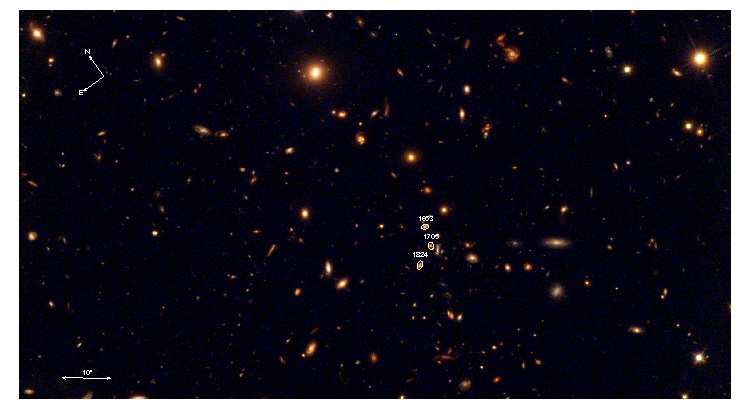}}
\caption[h]{Three color (F814W-F105W-F140W) image of the $z=1.40$ ISCS J1432.3+3253 cluster.
Red-sequence, early-type galaxies are marked. The late-type morphology of most
galaxies cannot be appreciated in this figure.
}
\end{figure*}

\begin{figure*}
\centerline{\includegraphics[width=18truecm]{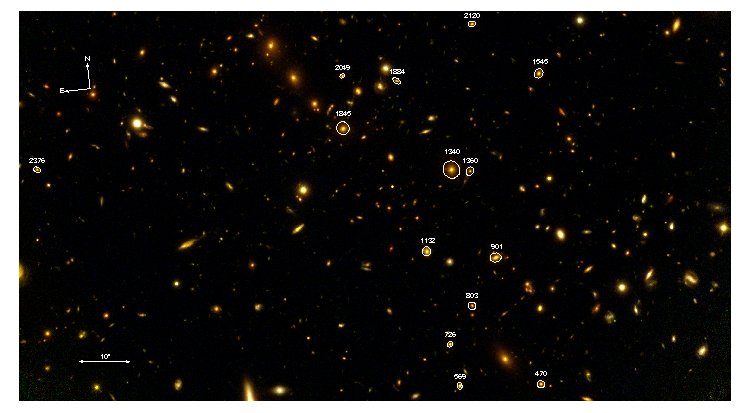}}
\caption[h]{Three color (F814W-F105W-F140W) image of the $z=1.32$ SPT-CL J0205-5829 cluster.
Red-sequence, early-type, galaxies are marked.  The late-type morphology of most
galaxies cannot be appreciated in this figure.
}
\end{figure*}

\section{Data}

\subsection{$z>1$ clusters}

The data used in this paper for $z>1$ clusters
are from the Wide Field Camera 3 near-infrared (NIR) and ultraviolet-visible 
(UVIS, Kimble et al. 2008) and 
Advanced Camera for Survey (ACS, Sirianni et al. 2005) 
wide field camera imaging of the following clusters: JKCS041 at $z=1.80$ 
(Andreon et al. 2009, Newman et al. 2014), IDCS J1426.5+3508 at $z=1.75$
(Stanford et al. 2012), SpARCS104922.6+564032.5 at $z=1.71$ (Webb et al. 2015), 
SpARCSJ022426-032330 at $z=1.63$ (Muzzin et al. 2013), 
XDCP J0044.0-2033 at $z=1.58$ (Santos et al. 2011), SPT-CL J2040-4451 at $z=1.48$
(Bayliss et al. 2014),
ISCS J1432.3+3253 at $z=1.40$ (Zeimann et al. 2013), and SPT-CL J0205-5829 at $z=1.32$
(Stalder et al. 2013). These are all $z>1.3$ clusters with HST observations appropriate
for this work (in depth and wavelength coverage) .
Details about exposure time and filters are
in Table 1. Most of the
data were observed for the various supernovae programs (PI Saul Perlmutter), many during
the preparation of this work.
We re-reduced all the data, about 166 HST orbits, 
starting from {\rm FLT} (near--infrared datasets) 
or {\rm FLC} (optical datasets) images to assure homogeneity between
cluster and control field observations. 
Images are combined and resampled to 0.06 arcsec pixels 
using Astrodrizzle (DrizzlePac 2.0.2.dev42994 for images
with $\lambda_c > 1 \mu m$,
or v1.1.16 otherwise).
Images in F105W are aligned with teakreg. These are used for alignment and
as reference images so that the other filters exactly match the pixel scale, 
center, and tangent point of the output mosaics. ACS images, corrected for
the effect of charge-transfer efficiency, were
aligned using the procedure described in Dong et al. (2015)
because there were too many cosmic rays. Briefly, 
for shift computation we use 
images with cosmic rays identified and flagged using L.A.COSMIC
(van Dokkum 2001), while during coadding we leave astrodrizzle 
to find and flag cosmic rays.

Figures 1 to 7 show HST three-color images of the seven clusters without an
earlier similar published image.

We use SExtractor (Bertin \& Arnouts 1996) for detection and photometry. We use
F140W for detection when it is availble (otherwise F160W), and we derive
photometry in the other bands running SExtractor in dual-imaging mode.
Colors are based on fluxes within the F140W 
isophote (when availble, otherwise F160W) with a minor
correction for PSF differences across filters. The correction is derived comparing
isophotal and
2.0 arcsec aperture colors of galaxies with isophotal sizes in the 
range of those studied in this paper ($0.7\lesssim r_{iso} <1.5$ arcsec). 

Only galaxies with $\log M/M_\odot \gtrsim 10.7$ (derived as described in \S 3)
and within $\pm 0.2$ mag of the color--magnitude relation
are considered, which corresponds to $F140W \sim 22.7$ mag, depending
on redshift. Images are at least two mag deeper than this limit, and also allow us
to characterize the faint end of the red sequence (Andreon et al., in prep.),
not just the bright galaxies studied in this work. Indeed, morphological
classification of galaxies in some of these clusters has already been performed in the past, sometimes
using a reduced exposure time (Newman et al. 2014; Stanford et al. 2012; Tracy et al. 2015)
or a worse resolution (Fassbender et al. 2014).

To estimate the number of back/foreground galaxies in the cluster line of sight
we use three fields observed with the same filter set used for clusters. 
They are the Hubble Ultra Deep Field (HUDF) and 
two parallel observations of the Frontier Fields. We note that 
the Frontier Fields parallels are overdense of galaxies at $z<0.6$
because they are parallel to intermediate redshift cluster observations. Contamination
in the Frontier Fields parallels can be easily recognized and removed because
these galaxies have a 4000 \AA \ break in a much bluer filter than the 
F105W used for color measurements and therefore cannot be
as red in $F105W-F140W$
as those of interest here.

\subsection{$z<1$ clusters}

To place our results in context and to have a more complete view of the redshift
evolution, we consider a cluster sample at $z<1$ with almost complete spectroscopic 
coverage so that the background contribution
can be identified and individually removed from the sample. These clusters, listed
in Table 2, are
analyzed as $z>1$ clusters. In particular, galaxies are
selected within $\pm 0.2$ mag of the red sequence using a color index bracketing
the 4000 \AA \ break.

RXJ0152.7-1357 ($z=0.84$) uses ACS images taken from 
the Hubble Legacy Archive with $0.05$ arcsec pixels. Spectroscopic redshifts 
are taken from Demarco et al. (2005, 2010). 

MACSJ1149.5+2223 ($z=0.544$) galaxies are initially photo-z member selected
(i.e., $|z_{phot}-z_{spec}|<3 \sigma_z$ where $\sigma_z=0.04*(1+z_{spec})$) 
using the Cluster Lensing and Supernova Survey with Hubble (CLASH, Postman et al. 2012) 
16-band photometric redshifts. We then keep galaxies
showing
a passive type (BPZ most likely spectral type $<5.5$). Finally,
galaxies are selected according to color and morphology.
Spectroscopic redshifts for almost
all galaxies of interest are available in Treu et al. (2015) 
and Ebeling et al. (2014), making the photo-z pre-selection
necessary only for excluding a few galaxies 
without a spectroscopic redshift but with a SED
pointing toward a manifestly different redshift. 
Similarly, the passive pre-selection is in principle unnecessary
because all non-passive galaxies are clearly non-early-type, which
can be determined by eye. 
For the morphological analysis we
use ACS images 
with $0.05$ arcsec pixels,
reduced in Andreon (2008). To test the effect of sampling and depth,
we also analyze a few galaxies using the $>20$ times longer exposures with $0.03$ 
arcsec pixels recently
acquired and available in the Hubble Legacy Archive.

MACSJ1206.2-0847 ($z=0.44$) galaxies are photo-z selected, as MACSJ1149.5+2223 ones
are, using CLASH (Postman et al. 2012) 16-band photometry, and then
galaxies are selected according to color and morphology.
Spectroscopic redshifts are available in Biviano et al. (2013), making 
the photo-z pre-selection almost un-necessary.
For the morphological analysis we
use ACS images 
with $0.065$ arcsec pixels from 
Postman et al. (2012).

Abell 2744 ($z=0.306$), also known as AC118, uses ACS images taken from 
the Hubble Legacy Archive with $0.03$ arcsec pixels. 
The F814W image is impressive for its
depth and sharpness. It allows a finer morphological analysis 
(e.g., detecting shells, lens, and external rings) of lower mass galaxies than needed here.
Spectroscopic redshifts for virtually
all galaxies of interest are available in Owers et al. (2011). 

Abell 2218 ($z=0.1773$) uses ACS images reduced for this work. Spectroscopic redshifts for all
but one galaxy of interest are available in Le Borgne et al. (1992),  
Sanchez et al. (2007), and Haines et al. (2015).

Abell 1656 (Coma) galaxies use ground-based $V-$ or $r-$band observations, typically 20 min
long with $<1.2$ arcsec seeing at a 2m telescope (from Andreon et al. 1996; 1997).
Effective radii, derived as in this work and using the same software,
are available in the references mentioned.

Overall, we reduced about 200 HST orbits for this paper (for the morphological
analysis of the high-redshift sample), and we analyzed
images for more of 300 HST orbits.

\section{Morphology, size, and stellar mass}

\begin{figure}
\centerline{\includegraphics[width=9truecm]{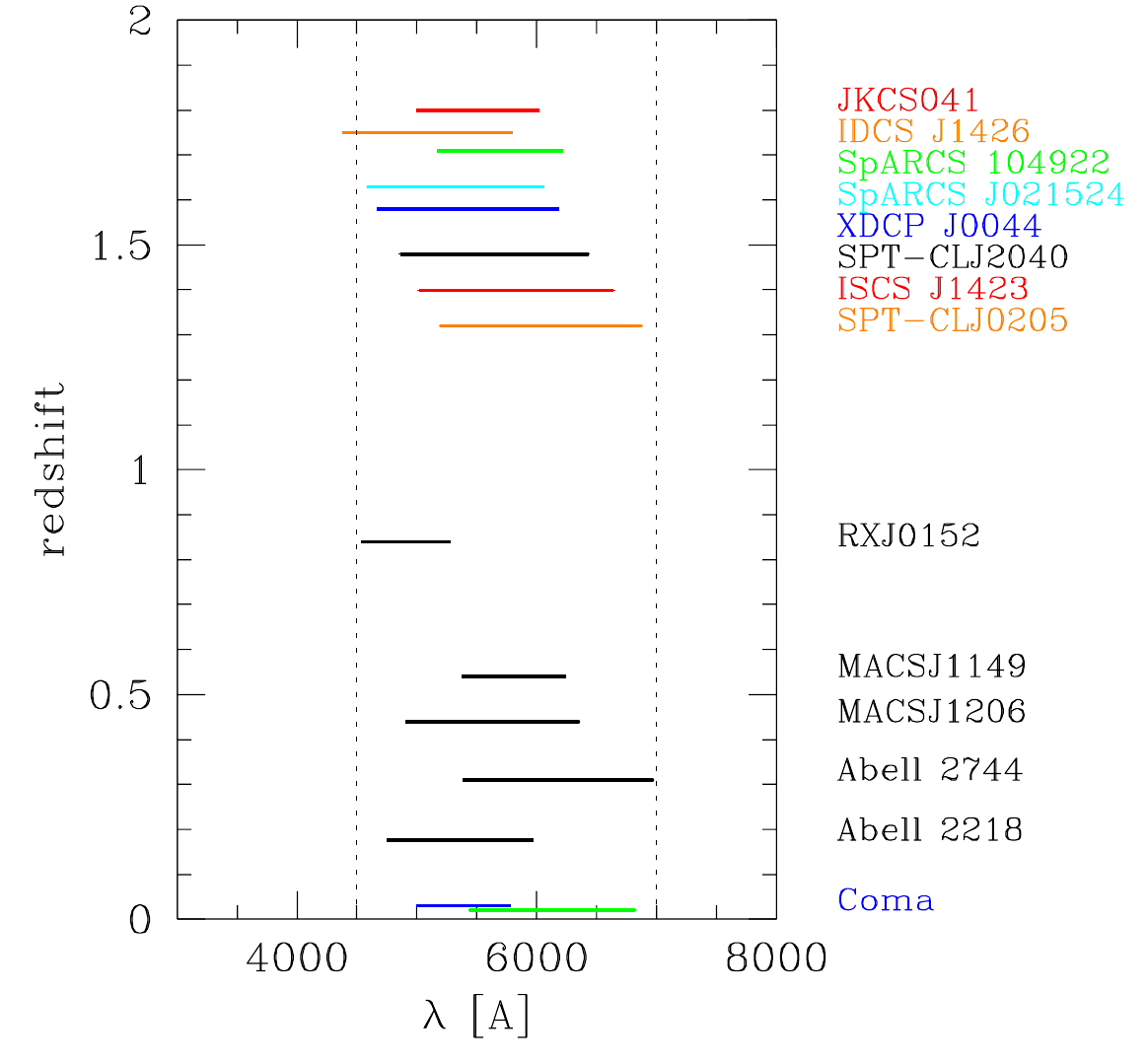}}
\caption[h]{Rest-frame wavelength sampling of the band used to 
derive morphologies, sizes, and masses. 
}
\end{figure}

In order to derive effective radii and total luminosities (to be used later for deriving
the galaxy mass) we fit the galaxy isophotes, 
precisely as done for galaxies in different environments at low and intermediate redshift
(e.g., Michard 1985, Poulain et al. 1992;
Michard \& Marshall 1993, 1994; Andreon 1994; Andreon et al. 1996, 1997, etc).

Isophotes are decomposed in ellipses plus Fourier coefficients
(Carter et al. 1978, Bender \& Moellenhoff 1987, Michard \& Simien 1988)
to describe deviations from the perfect elliptical shape. Each isophote
has a center, major and minor axes, position angle, and Fourier coefficients measuring
deviations from a perfect ellipse. Parameters can differ from isophote to isophote
allowing us to describe the brightness distribution of galaxies with
structural components such as bars, disks, bulges,
spiral arms, and HII regions. These structural components have distinctive
signatures in the radial profiles of the isophote parameters, for
example bars shows up as changes in position angle and/or axis ratio,
disks are associated with changes of ellipticity (except when
face on), spiral arms and other irregularities in the isophote shapes 
are measured by (some) non-zero Fourier coefficients, as
shown in the papers mentioned above and also in Peng et al. (2010). 
We therefore 
classify galaxies by detecting morphological components in the
radial profiles of the isophote parameters. Such a classification,
based on measurement of distinctive features of morphological
components, returns morphologies coincident
with the ones performed by morphologists such as Hubble, Sandage, 
de Vaucouleurs, and Dressler (Michard \& Marshall 1994; 
Andreon \& Davoust 1997) who
use visual inspection, rather than measurements, 
to detect structural components. Based on measurements,
structural classification is more reproducible than morphologies
based on visual inspection (Andreon \& Davoust 1997).

This morphological classification allows us to
remove from the sample non-early-type galaxies (i.e., spirals and irregulars),
only keeping early-type ellipical and lenticular galaxies.
Visual inspection of each of them in the different bands always confirmed 
the non-early-type nature of these galaxies, provided that
images are displayed with the appropriate cuts, which is sometime
cumbersome\footnote{In turn, our structural analysis can be
used to qualify the quality of the stacked images: the
morphological components leaves such standard signatures in the isophote parameter
radial profile that we are able to discover
from the unusual shape of ther radial profiles of Fourier terms 
that the stacked MACSJ1206-08 F606W image delivered by CLASH 
(Postman et al. 2012) 
has been not properly combined.
The HST archive confirmed us that the combined images have different
distortion patterns as a result of the large time baseline over which
they were taken, but only one single distortion pattern has been
used to produce the stack.}.  

To compute the total galaxy flux, and from it the galaxy mass and size,
the flux between isophotes is analytically integrated up to the last detected isophote,
hence determining the curve of growth.
To extrapolate it to infinity, we fit the measured growth curve
with a library of growth curves measured for galaxies of different morphological
types in the nearby universe (de Vaucouleurs 1997), keeping the one that fits best.
The half-light isophote is, by definition, the isophote including half the 
total light. The half-light circularized radius, $r_e$, is defined 
as the square root of area included in the half-light
isophote divided by $\pi$. This definition allows us to 
define the half--light radius whatever the isophote shapes are and irrespective of 
wheter galaxies has a single value of ellipticity and position angle, or
values that depend on radius,
as barred galaxies, lenticulars, and many ellipticals have.
The background light is accounted for, and subtracted, by
fitting a low-order polynomial to the region surrounding the studied galaxy,
and accounting for the galaxy flux at large radii.
This also allows us to remove any
residual gradient present in the image, that is due to, e.g.,
scattered or intracluster light.

Half-light radii derived from the curve of growth have been extensively used in
the literature (e.g., Sandage 1972, de Vaucouleurs, de Vaucouleurs, and Corwin 1976; 
Davies et al. 1983;
Burstein et al. 1987; Dressler et al. 1987, Lucey et al. 1991; Bender et al. 1992;
Saglia et al. 1993a; Jorgensen et al. 1995; Giavalisco et al. 1996;
Andreon et al. 1996, 1997; Prugniel \& Simien 1996;
Pahre et al. 1995, 1998; Saglia et al. 2010) 
and the derivations by the different astronomers have been thoroughly and extensively
compared (see the papers above and references therein). 
In particular, size derivation using the curve of growth, allowing the observed variety
of morphological structures in galaxies, and HST data have been used
since the HST repair (Andreon et al. 1997).

Masses of red-sequence early-type
galaxies are derived from $\lambda \approx 6000$ \AA \  luminosities 
assuming our standard BC03 SSP model with $z_f=3$ (which in turn
matches the red-sequence color). For
red-sequence galaxies in JKCS041
this has been shown to introduce a negligible $0.10$ dex
scatter in mass and no bias
compared to a derivation based on fitting 12 bands photometry
and $3000-6000$ \AA \ spectroscopy (Andreon et al. 2014). 
A further check is given in Sec.~4.
Fig.~8 shows the uniform rest-frame sampling
of the band used to derive morphologies, sizes, and masses.

\begin{figure*}
\centerline{\includegraphics[width=12truecm]{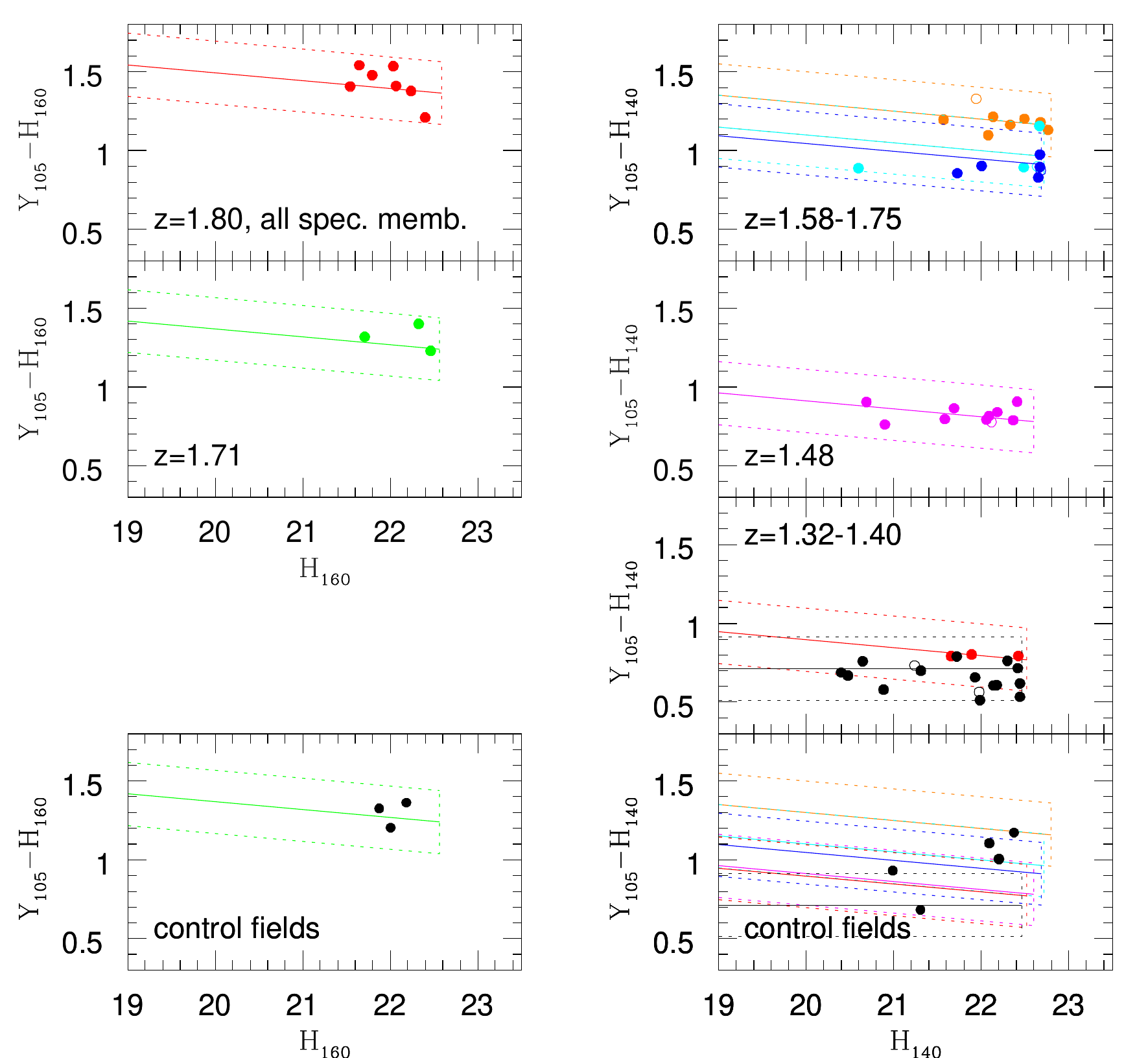}}
\caption[h]{Color--magnitude plot of red-sequence early-type galaxies. The slanted rectangles
indicate the selection region ($\pm 0.2$ mag from the color-magnitude relation),
with $H$ magnitude brighter than a SSP with $\log M/M_{\odot}=10.7$, both
in the cluster (upper four panels) and reference (bottom panels) lines of sight. 
Clusters are color-coded as in Fig.~8. The open points
indicate galaxies within the boundaries but with unfeasible isophotal analysis.
}
\end{figure*}

\section{Results}

Table 3 lists coordinates, mass, and size (half-light radius) of 158 
red-sequence galaxies studied in this work. 86 more are listed in
Andreon et al. (1996, 1997).
Table 1 and 2 summarize how many early-type galaxies more massive than $\log M/M_\odot >10.7$
there are in each cluster. Except Coma, with 86
galaxies, each cluster has between 3 and 23 red-sequence early-type 
galaxies.

Figure~9 shows the color-magnitude relation of red-sequence
early-type galaxies for clusters at $z>1$. 
The region from which red-sequence
early-type galaxies are selected ($\pm 0.2$ mag around the color-magnitude
relation and $\log M/M_\odot >10.7$) is shown in Fig.~9, both in
the cluster and control field panels. There are only few galaxies in
the control fields, and an even lower number is expected
in the cluster line of sight because the 
solid angle of the latter is about three time smaller.
From control field observations we computed
the expected number of back/foreground galaxies 
in the 
cluster line of sight in the appropriate region of the color-magnitude
plane (Table~1). We expect that out of 45 galaxies in the seven clusters at $1.3<z<1.79$,
only 5.5 are background galaxies. Therefore, 
the background contamination is minor ($\sim 10\%$)
and a negligible source of error.
We therefore neglect the background subtraction. 
JKCS041 has exactly zero background galaxies because 
all red-sequence early-type galaxies considered
in this work are spectroscopic members (Newman et al. 2014).

Very few galaxies, less than one per cluster on average, 
are in such crowded environments that they cannot be
reliably analyzed because their isophotes are, within a range of brightnesses,
too much contaminated by the presence
of other galaxies. These few galaxies, shown as open symbols in Figure~9, 
are therefore ignored in this work.

\begin{center}
\begin{table}
\caption{Coordinates, masses, sizes, and PSF corrections.}
\begin{tabular}{l r r r r r}
\hline
\hline
Id & R.A. & Dec. & $\log M/M_\odot$ & $\log r_e$ & PSF corr \\
 & \multispan{2}{\hfill J2000 \hfill} & & [kpc] \\
\hline
\multispan{5}{JKCS041} \\ 
2045 & 36.67527 & -4.70738 & 10.88 & 0.02 & -0.10 \\ 
982 & 36.68790 & -4.68994 & 11.06 & 0.15 & -0.06 \\ 
988 & 36.69051 & -4.69215 & 11.04 & 0.15 & -0.06 \\ 
...\\
\hline		   
\multispan{5}{Abell 2218 \hfill}\\ 
...\\
4286 & 248.97871 & 66.18289 & 10.68 & 0.11 &  0.00 \\ 
1606 & 248.91637 & 66.21551 & 10.66 & 0.13 &  0.00 \\ 
721 & 249.01959 & 66.22836 & 10.76 & 0.17 &  0.00 \\ 
\hline  	  
\hline
\end{tabular}					 
\hfill \break
Table 3 is enterely available in electronic form
at the CDS via anonymous ftp to cdsarc.u-strasbg.fr (130.79.128.5)
or via http://cdsweb.u-strasbg.fr/cgi-bin/qcat?J/A+A/
More digits than needed are reported for all quantities.
\hfill \break	  
\end{table}
\end{center}

Appendix A provides comments on individual galaxies or clusters. 
With very few, if any, exceptions, galaxies  that are are morphologically late 
and on the red sequence in the selection color and are blue in bluer 
color indexes. Furthermore, at $z<1$ about 83\%  of
red-sequence galaxies have a known spectroscopic membership.

Morphologies, effective radii, and masses derived for seven 
(random) red-sequence early-type galaxies
in MACSJ1149.5+2223, independently derived from images taken at different epochs
(more than 10 yrs away), having widely different depths and
sampling, and that have been stacked by different authors (Andreon 2008 and
the Frontier Field initiative) are consistently the same, showing that  
improvements in the image alignment and stacking, a 20-fold increase in
exposure time, and the additional resolution of a smaller pixel size are inconsequential 
for the bright objects studied in this work. Quantitatively, the morphologies
are identical, the scatter in size is $\sigma(\log r_e)=0.06$ dex, and the scatter in
mass is $\sigma(\log M/M_\odot)=0.04$ dex.

Our simple recipe for mass computation has been tested with the galaxies in RXJ0152.7-1357,
whose masses have been derived in Delaye et al. (2014) from SED fitting.
The mass-size relation of this cluster (derived in \S 4.2),
based on masses derived from our recipe, turns out to be identical to
the one using masses derived from SED fitting, confirming the
robustness of deriving masses for red-sequence early-type
galaxies from a $\lambda \sim 6000$ \AA \ total magnitude.

\subsection{Smaller size at high redshift}

The top panel of 
Figure~10 shows the mass-size relation of red-sequence early-type galaxies color-coded
by cluster ID for clusters at $z>1.45$. As a comparison, the bottom panel
shows Coma galaxies, which occupy the wedge between the two
slanted lines, also shown in the top panel. 
As can be seen, at a given mass only the less extended galaxies
are present in $z>1.45$ clusters. The right panel of 
Figure~11 quantifies this by
plotting the size distribution, reduced to $\log M/M_{\odot}=11$ 
(i.e., corrected for the mass-size relation) for the Coma and
our combined $z>1.45$ cluster sample, i.e., the projection of Fig.~10 along a
line of slope $0.6$. This quantity is sometimes referred to as mass-normalized size (Newman
et al. 2012; Cimatti et al. 2012; Delaye et al. 2014). Accounting for
the slope between size and mass is important 
in order to reduce the effect of potentially different
mass distributions across the samples, i.e., to discriminate real differences in size at a
given mass from those spuriously induced by differences in mass across the samples.
To match sample sizes, we
normalize the Coma histogram by matching the number of galaxies with $10.7<\log M/M_{\odot}<10.9$
(see also the left-hand panel). 
As can be seen from Fig.~10 and from the right--hand panels of Fig.~11, Coma galaxies 
are on average larger than our high-redshift sample: there are
large ($>3$ kpc) galaxies in Coma, which are absent in high-redshift clusters, 
and there is an excess of small ($<2$ kpc) galaxies in the  high-redshift 
sample, the smallest of which are absent in the low-redshift sample. 
A Kolmogorov-Smirnov test gives
a vanishingly small ($\sim 4 \ 10^{-4}$) ``probability''\footnote{To be precise,
this is a p-value, not a probability.}.
The mass distributions of the two
samples are shown in the left-hand panel of Fig.~11.
They are quite similar (p-value: $0.13$). Similar results are obtained replacing
Coma with the two clusters at lower redshift observed with HST, indicating
that the found size growth is not due to a Coma peculiarity.

We have so far neglected the effect of PSF blurring on half-radii
to show that differences are not caused by the adoption of possibly incorrect 
point spread function (PSF) corrections. PSF blurring make
high-redshift galaxies apparently 
larger, making even more significant the smaller size found at
high redshift.

\begin{figure}
\centerline{\includegraphics[width=9truecm]{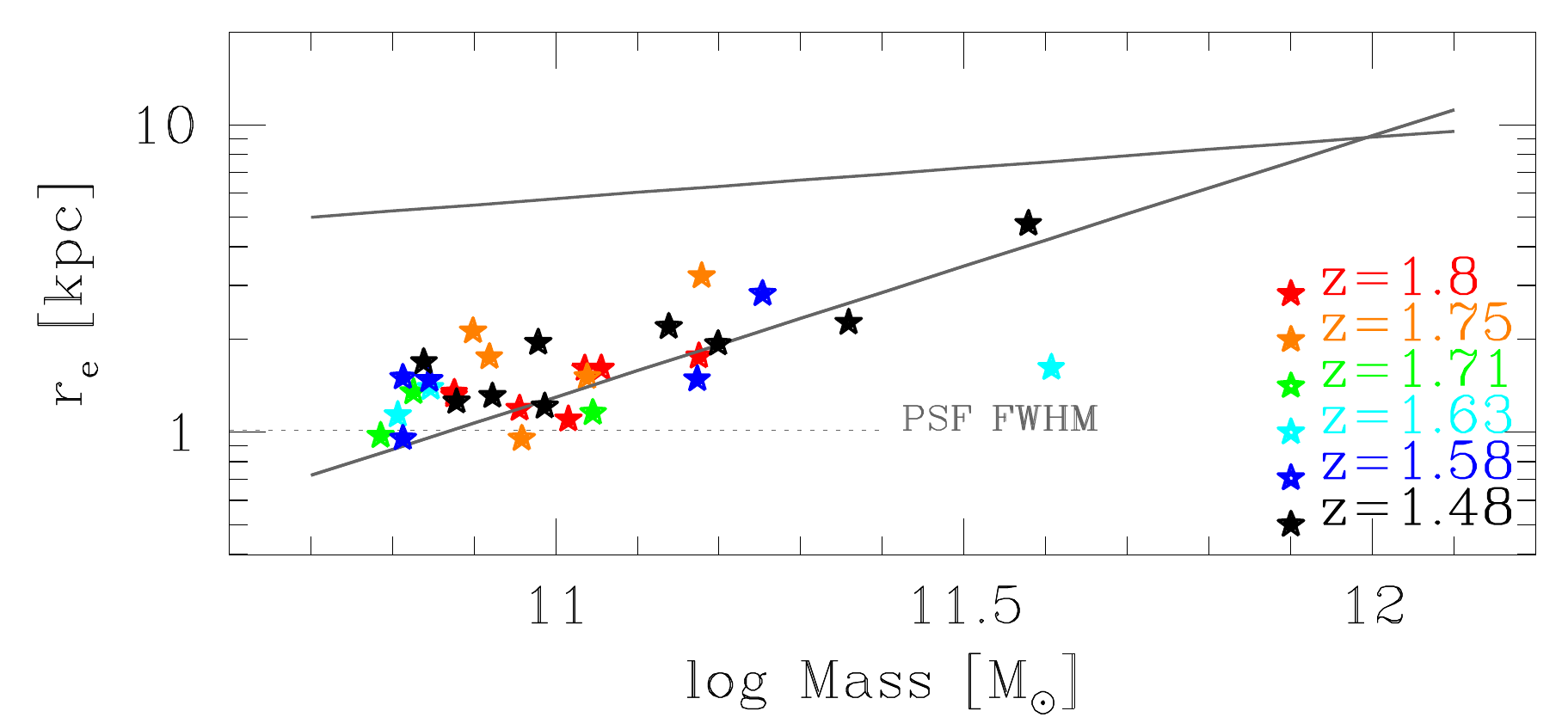}}
\centerline{\includegraphics[width=9truecm]{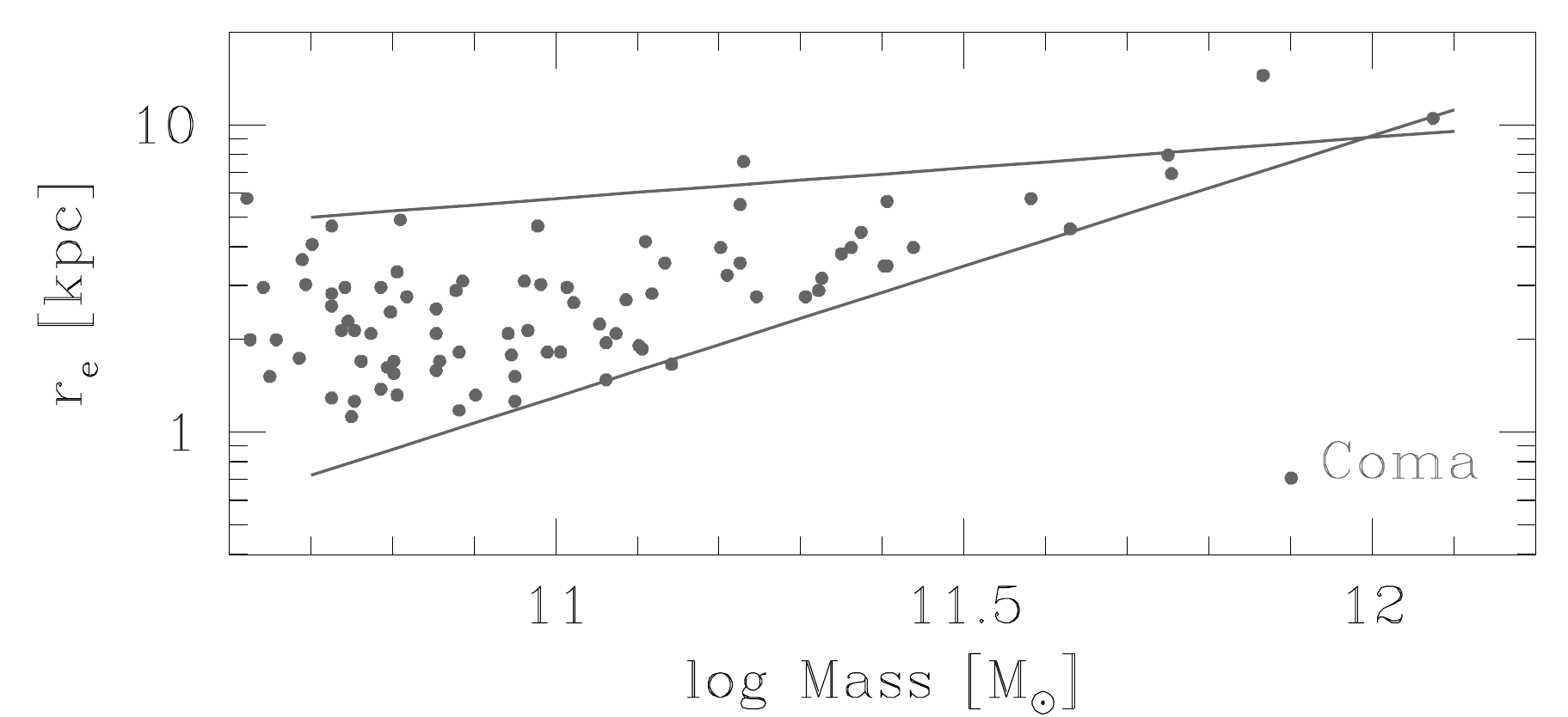}}
\caption[h]{Mass-size relation of red-sequence early-type cluster galaxies
at $z>1.47$ (points color-coded, upper panel) or Coma ($z=0.00232$, bottom
panel). Coma cluster galaxies occupy
the wedge between the two slanted lines, reported in both panels. 
The horizontal dotted line in the top panel indicates
the PSF FWHM. Sizes are not corrected for PSF blurring effects.
}
\end{figure}

\begin{figure}
\centerline{\includegraphics[width=9truecm]{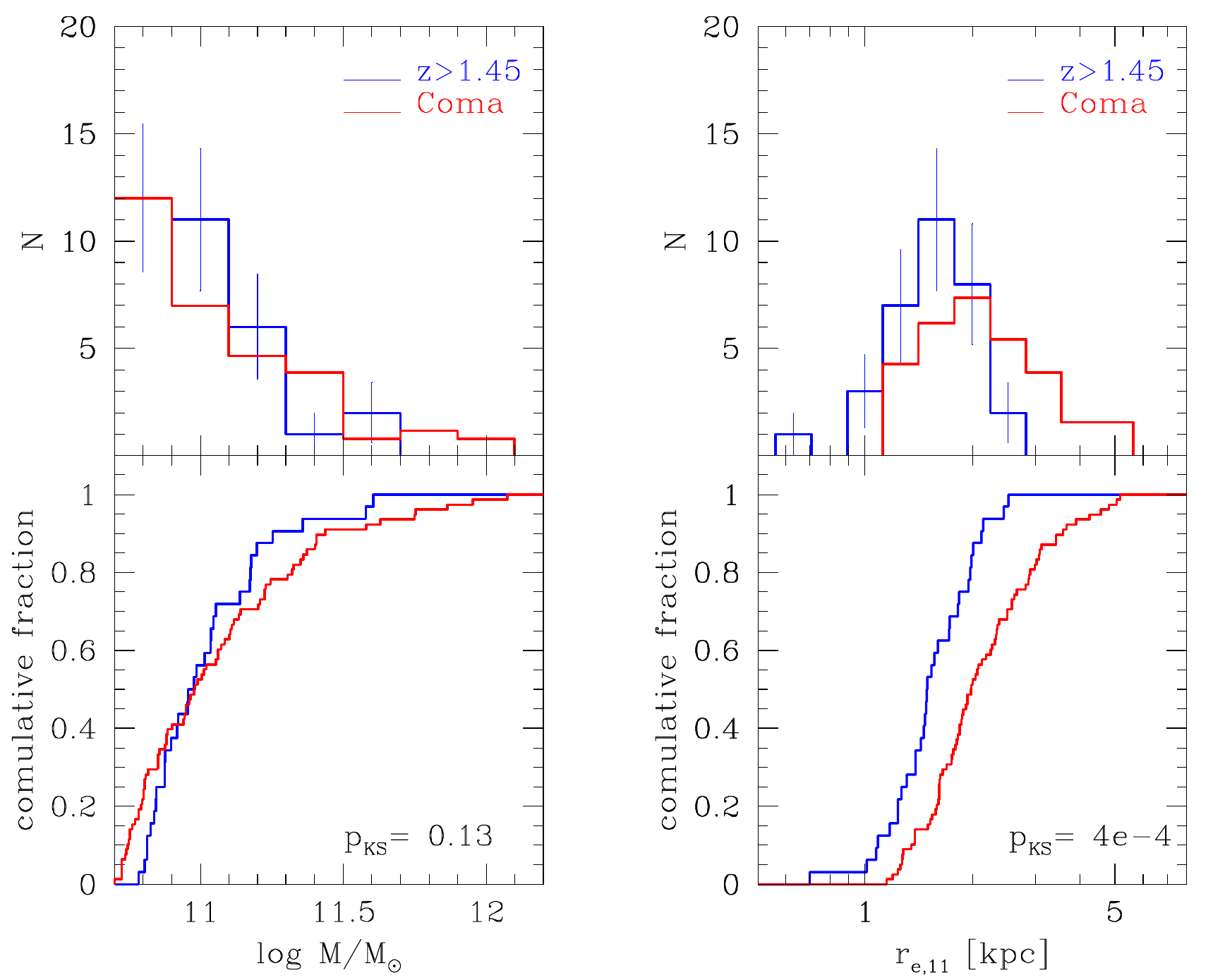}}
\caption[h]{Mass distribution (left) and size distribution (right) 
of $\log M/M_{\odot}>10.7$ galaxies of high-redshift (blue)
or Coma (red) cluster galaxies.  Sizes are not corrected for PSF blurring effects.
To account for the different richness of
compared samples, we matched the number of galaxies with masses in the lowest
mass bin. The outlier with smallest effective radius reduced to $\log M/M_{\odot}=11$ 
is the peculiar BCG of the $z=1.63$ cluster. 
}
\end{figure}

\begin{figure}
\centerline{\includegraphics[width=9truecm]{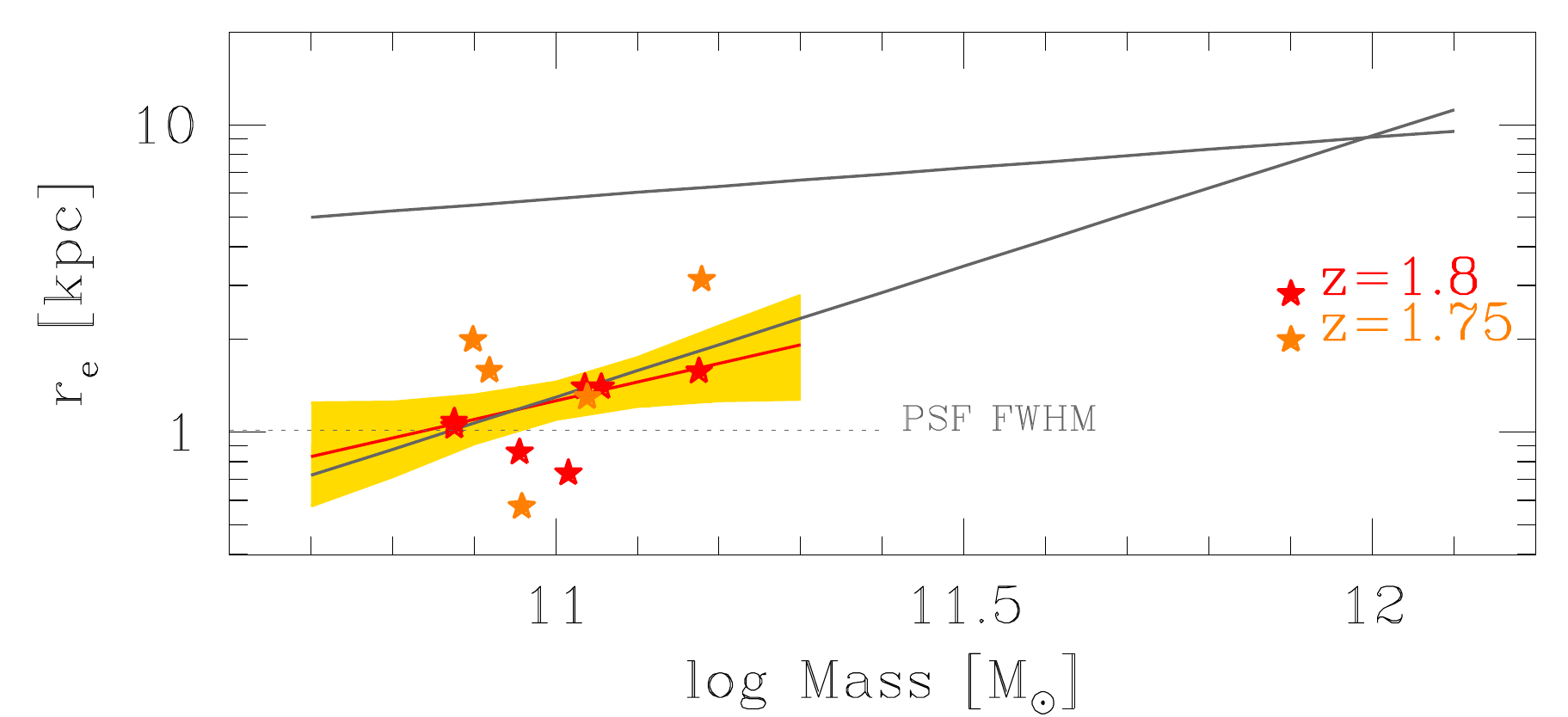}}
\centerline{\includegraphics[width=9truecm]{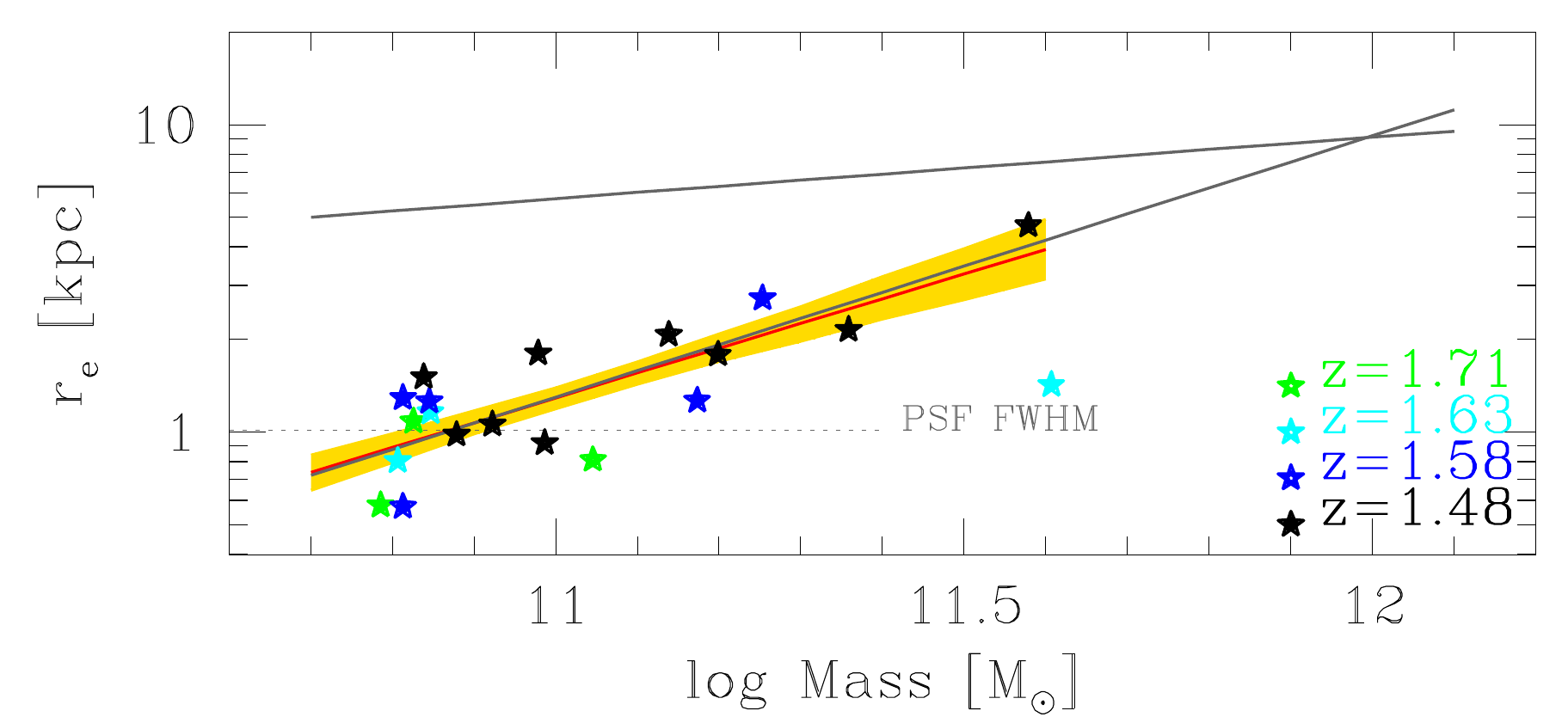}}
\centerline{\includegraphics[width=9truecm]{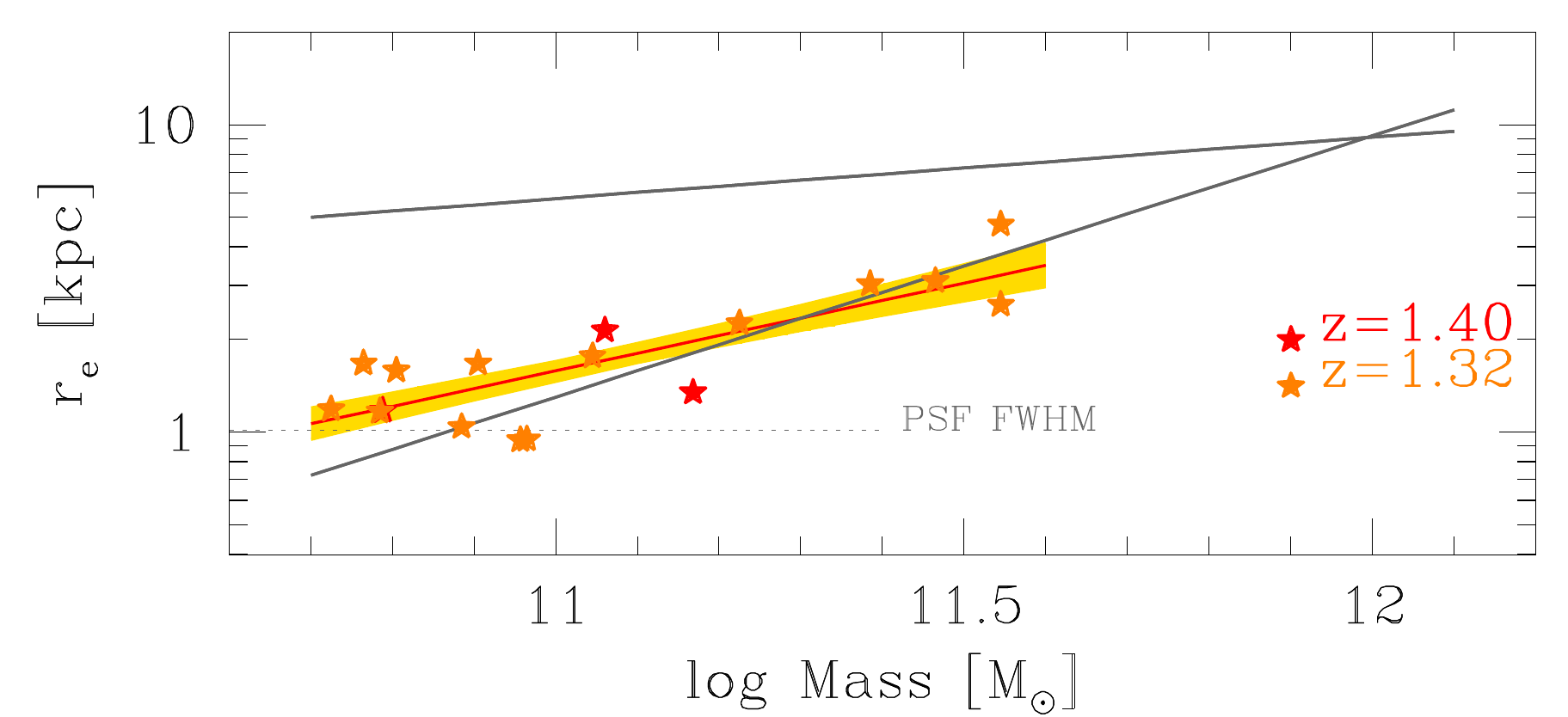}}
\caption[h]{Mass-size relation of red-sequence early-type cluster galaxies 
at $z>1$.  Sizes are corrected for PSF blurring effects.
The solid line and shading shows the fitted mass-size
relation and its 68 \% uncertainty (posterior highest density interval).
The wedge between the two slanted lines is the locus of Coma galaxies.
The horizontal dotted line indicates the PSF FWHM.
}
\end{figure}

\begin{figure}
\centerline{\includegraphics[width=9truecm]{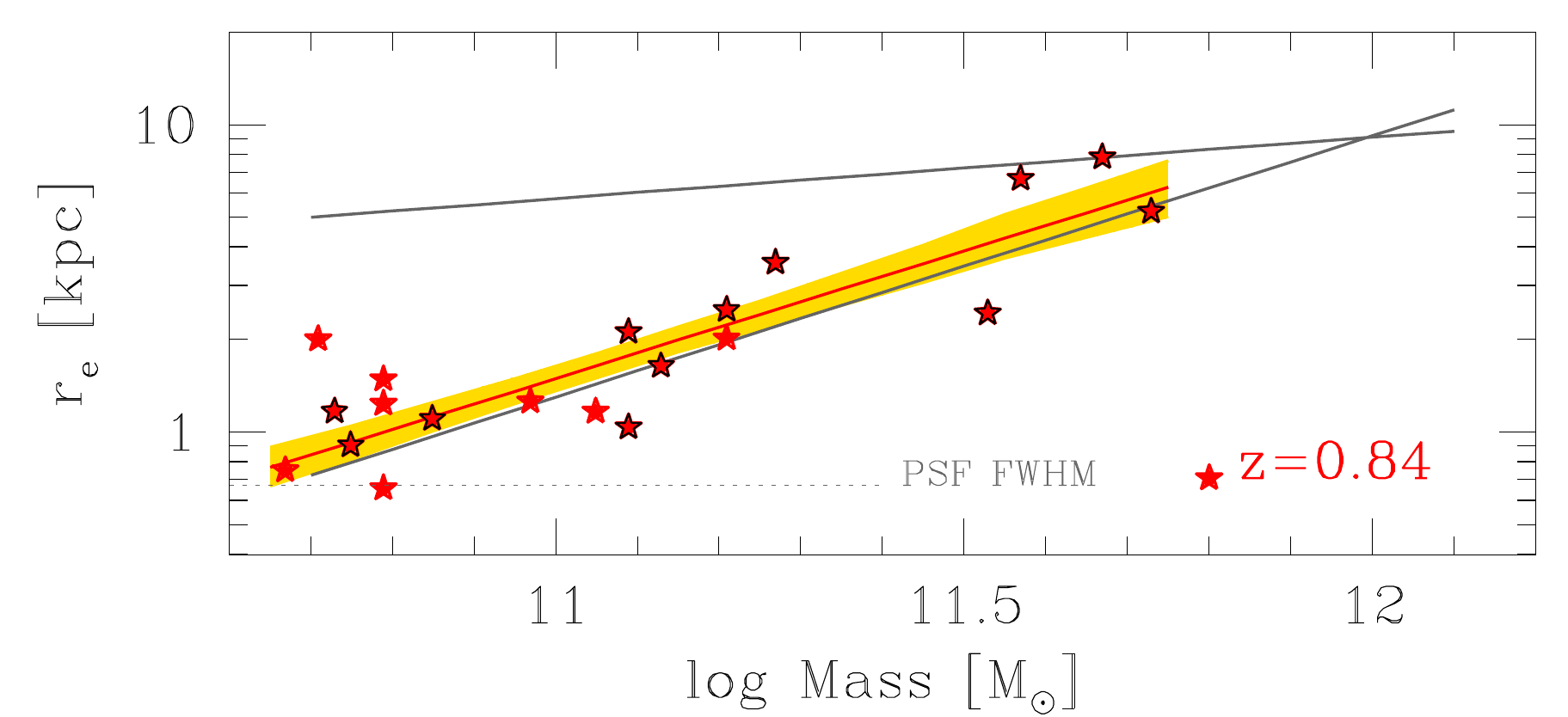}}
\centerline{\includegraphics[width=9truecm]{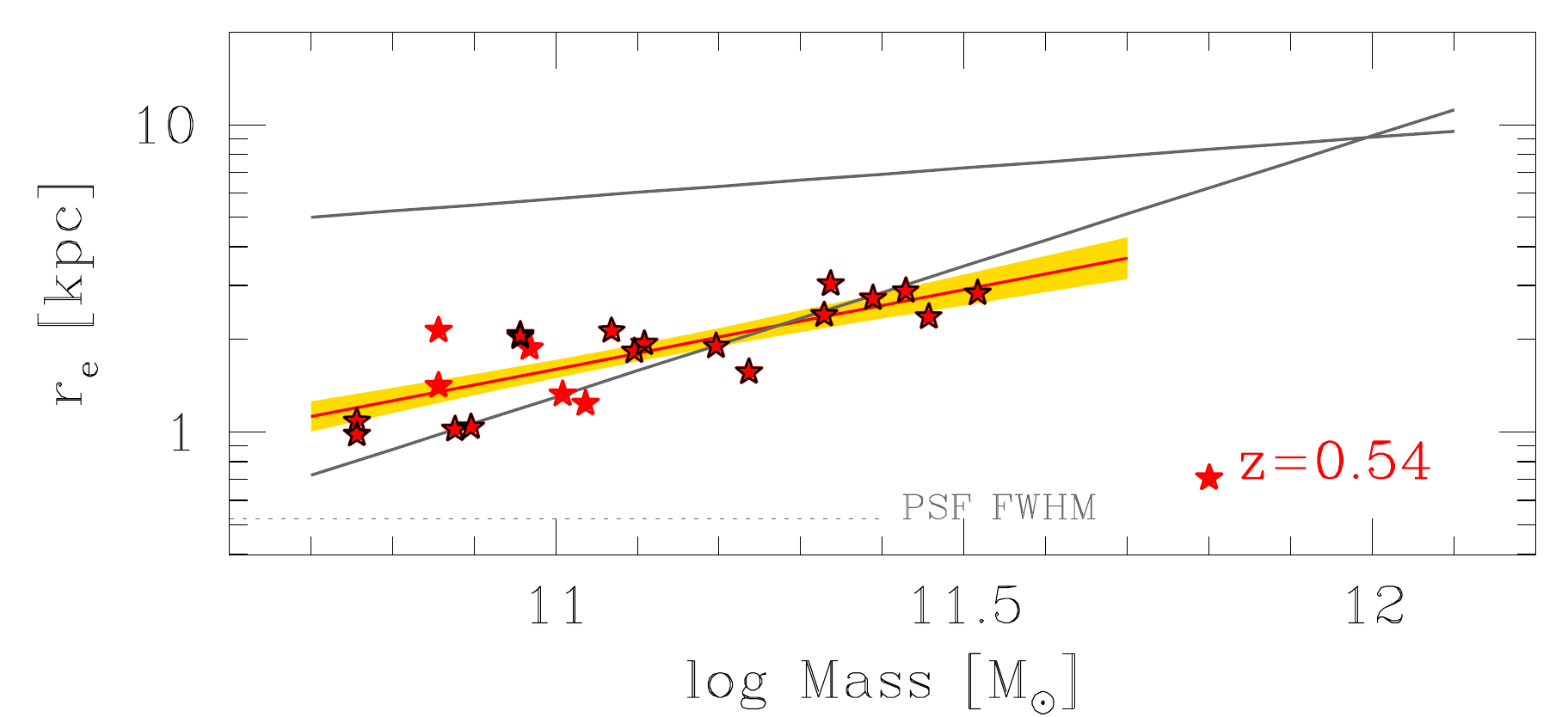}}
\centerline{\includegraphics[width=9truecm]{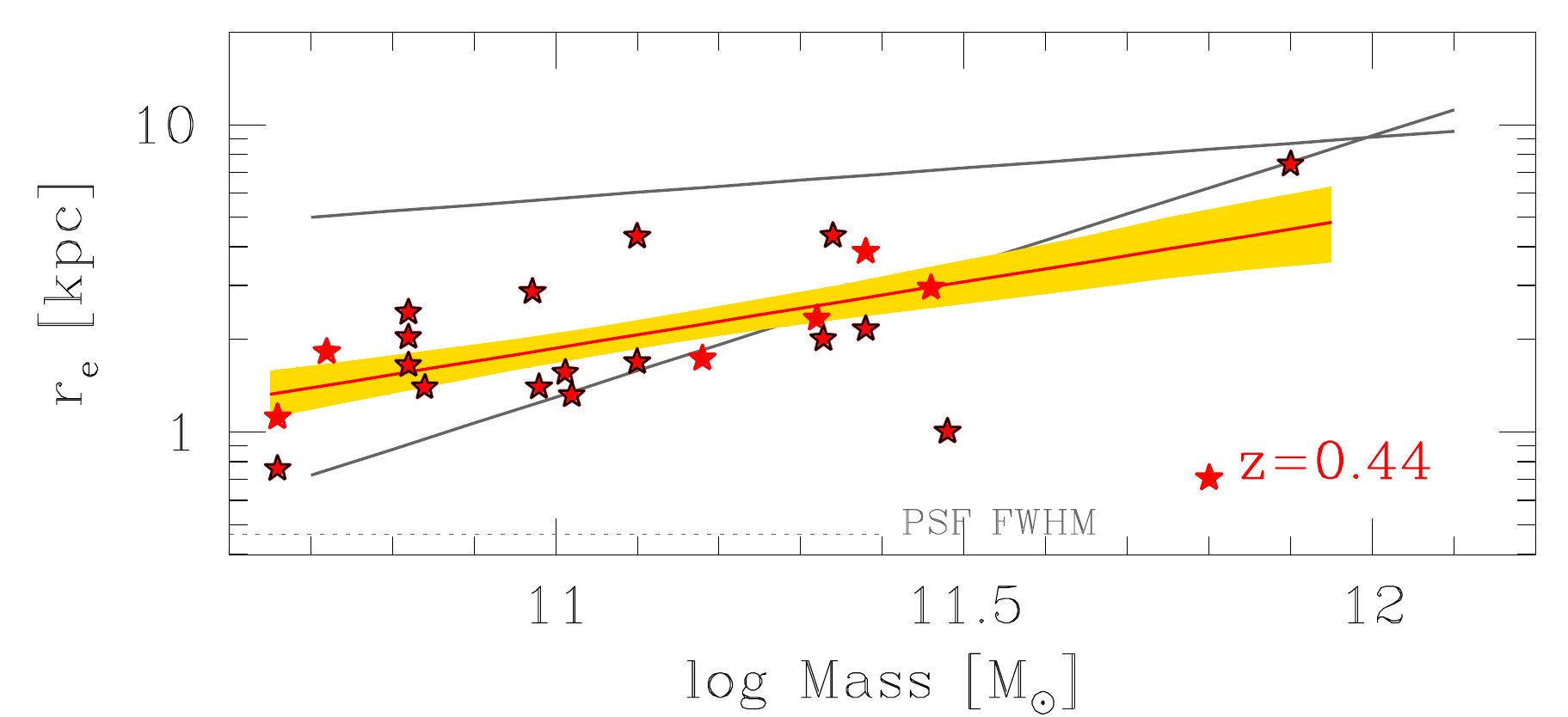}}
\centerline{\includegraphics[width=9truecm]{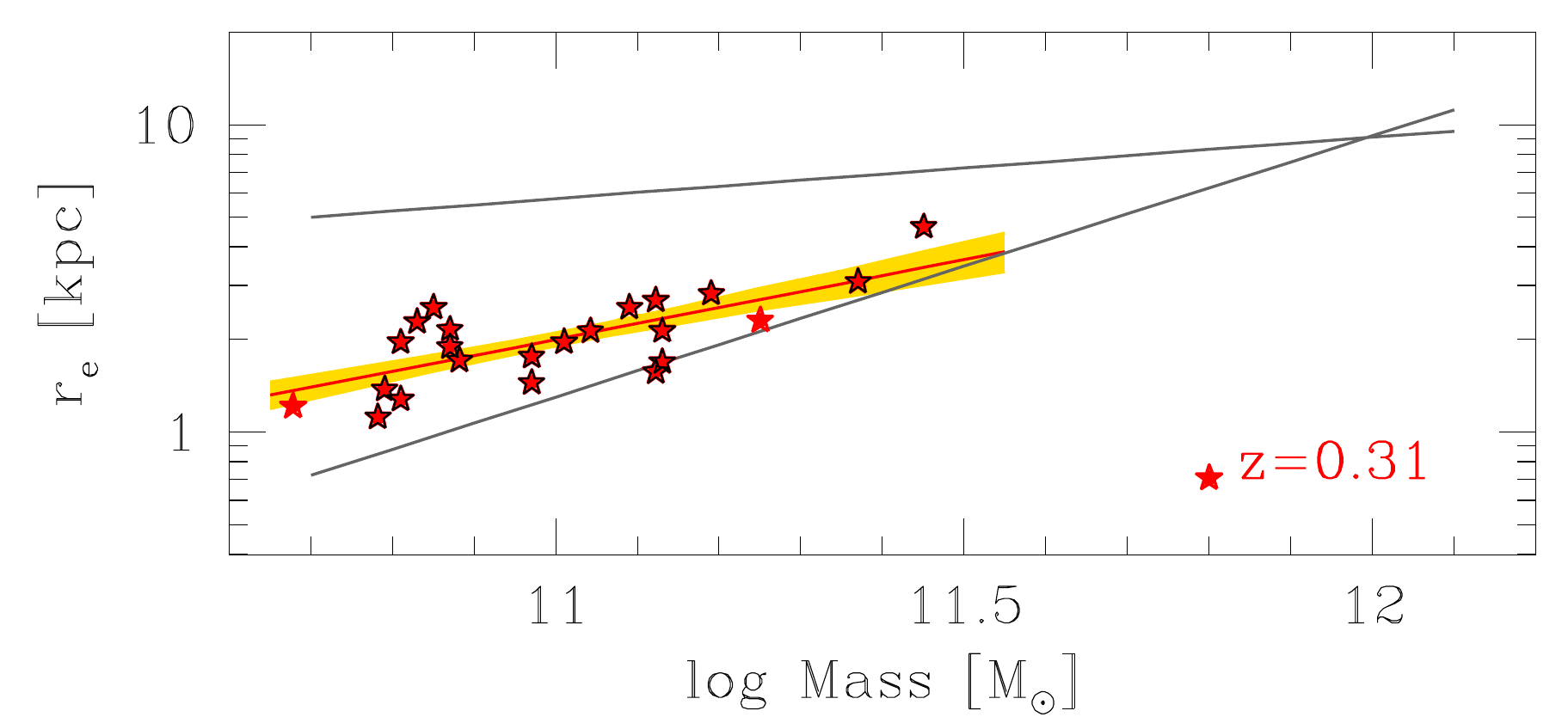}}
\centerline{\includegraphics[width=9truecm]{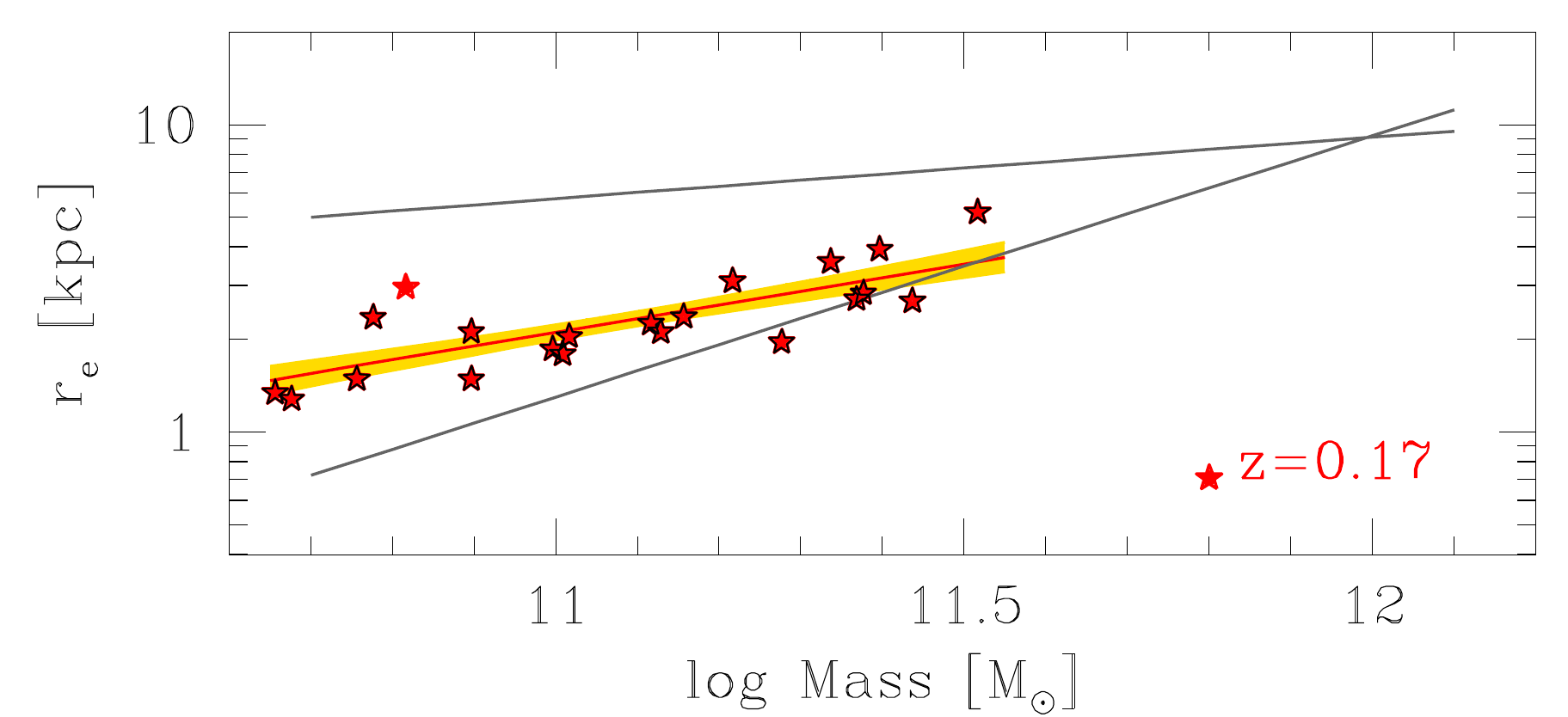}}
\caption[h]{Mass-size relation of red-sequence early-type cluster galaxies
at $z<1$. From top to bottom: RXJ0152.7-1357 ($z=0.84$),
MACSJ1149.5+2223 ($z=0.544$), MACSJ1206.2-0847 ($z=0.44$), 
Abell 2744 ($z=0.31$), and Abell 2218 ($z=0.17$).
The relation for Coma is shown in the bottom
panel of Fig.~10. Sizes are corrected for PSF blurring effects,
but the correction is so small that it cannot be seen for these
clusters.
Points with black contours are spectroscopically confirmed galaxies.
The red solid line and yellow shading show the fitted mass-size
relation and its 68 \% uncertainty (posterior highest density interval).
The wedge between the two slanted lines is the locus of Coma galaxies.
The horizontal dotted line indicates the PSF FWHM (for Abell 2744 and
Abell 2217 it is smaller than the displayed $r_e$ range).
}
\end{figure}

\subsection{PSF effects}

The PSF smears images and therefore makes galaxies to appear larger
than they actually are.
We correct for PSF blurring by computing, following Saglia et al. (1993b),
the size correction as a function of the observed half-light radius
expressed in FWHM units and assuming an $r^{1/4}$ radial profile. 
We applied the correction on a galaxy-by-galaxy basis, and we list
the applied correction in Table~3.
The correction is, in practice, zero at $z<0.8$, negligible at $z=0.84$,
and then increases at higher redshifts mostly because of the broader PSF in NIR.
The corrections remain small even at high redshift 
(for the two highest redshift clusters it has mean $-0.08$ dex and median $-0.06$ dex)
basically because the average galaxy with $\log M/M_{\odot}=11$ has 
$r_e > FWHM$, where the correction is small.
The NIR channel PSFs are approximated well by Gaussian profiles (Dressel 2012),
and therefore these profiles are assumed in our calculations.
From now on, all size measurements are
corrected for PSF blurring.

\subsection{How did the mass-size relation evolve?}

In this section we determine how galaxy sizes (half-light radii) evolved by monitoring the
mass-size relation at different redshifts. We fit the mass-size relation 
using a linear model with
intrinsic scatter $\sigma$ of the form
\begin{equation}
\log r_e = \gamma \ +\alpha (\log M/M_\odot -11)
\end{equation}
adopting uniform priors  
for all parameters except the slope, for which we took instead a uniform prior 
on the angle. The parameter $\gamma$ is, by definition, the average size at 
$\log M/M_{\odot}=11$.  
By allowing a non-zero scatter, the information
content of each individual point has a minimal floor given by the large scatter, 
rather than by the smaller uncertainty of each size determination. 
Our approach, therefore, improves upon
those works that only consider size errors as the unique source of uncertainty.
By allowing the slope to be free, we de-weight
galaxies with masses fairly different from $\log M/M_{\odot}=11$ in the determination
of the average size at $\log M/M_{\odot}=11$ (i.e., $\gamma$), hence improving
upon some previous analyses that hold the slope fixed, for example,
Carollo et al. (2013) who assume $\alpha=0$ or Yano et al. (2016) who
assume $\alpha=0.75$. By leaving the slope free
we also allow different evolutions for galaxies of different mass,
discarded a priori by those works that keep the slope fixed. 
We use, as mentioned, PSF corrected sizes. Because
PSF corrections are small, our analysis is robust
to PSF corrections and minimally changed if these are neglected.

Fig.~12 and 13 show the mass-size relation for the various samples: 
the data, the fitted trend and its uncertainty. Fit parameters are listed in 
Table~4. Although not strictly necessary, we
combine clusters at adjacent redshifts to reach a minimal number of $\sim20$
galaxies per fit. 

Fig.~14 shows the effective radius at $\log M/M_\odot =11$ (i.e., $\gamma$) as
a function of time. We fitted a linear relation adopting uniform priors for 
the intercept and the angle. We found
\begin{equation}
\log r_{e,11} (t) = 0.36 \pm 0.01 +(0.023\pm 0.002)(t-13.5) \ .
\end{equation}
The mean size of red-sequence early-type galaxies 
has grown by $5.4\pm0.5$ \% per Gyr at a fixed stellar mass over the last 10 Gyr.
In other words, 10 Gyr ago sizes were 58 \% of the present-day sizes, or
today the sizes are 1.7 times those of 10 Gyr ago.
PSF corrections have little effect on the result: by neglecting them we would
have found a slighly smaller evolution (about 3\%) because of the overestimated
sized at high redshift.  The
effective radius at $z=0$ and $\log M/M_\odot =11$ is 2.1 kpc with
4\% error, which is also independently confirmed using effective radii from other deep and 
high-resolution images of Coma galaxies in Jorgensen et al. (1995, 1999; plotted in Figure~14),  
and other authors as well (e.g., Saglia et al. 1993a, Aguerri et al. 2004, Hoyos et al. 2011, 
details in Andreon et al. 2014). 
Fig.~14 also shows the evolution determined
by Newman et al. (2012) for field galaxies, which is much larger over the same
redshift range. This curve should not be over-interpreted; it only gives the
relative size evolution. The absolute normalization is arbitrary and
depends on a number of things, such as the adopted initial stellar mass function, the
way size is measured, and how the population under study is selected.

The scatter in size at a given mass does not show any clear trend with Universe age,
staying constant at $0.1-0.2$ dex. From inspection of
Fig.~10 it is tempting to infer that the observed evolution
in size at a given mass is produced by a new population occupying 
the upper part of the wedge at low redshift, i.e., that new low-redshift
galaxies have large sizes at a
given mass. However, we find no evidence of this effect
with the current dataset after accounting for the evolution
of the mean size at a given mass. 

To summarize, we find that the $\log$ of the galaxy size at a fixed stellar
mass increases
with time with a minor rate in the last 10 Gyr,
in marked contrast with the larger increase found in the literature
for galaxies in the general field over
the same period (see Sect.~5.1).

\begin{figure}
\centerline{\includegraphics[width=9truecm]{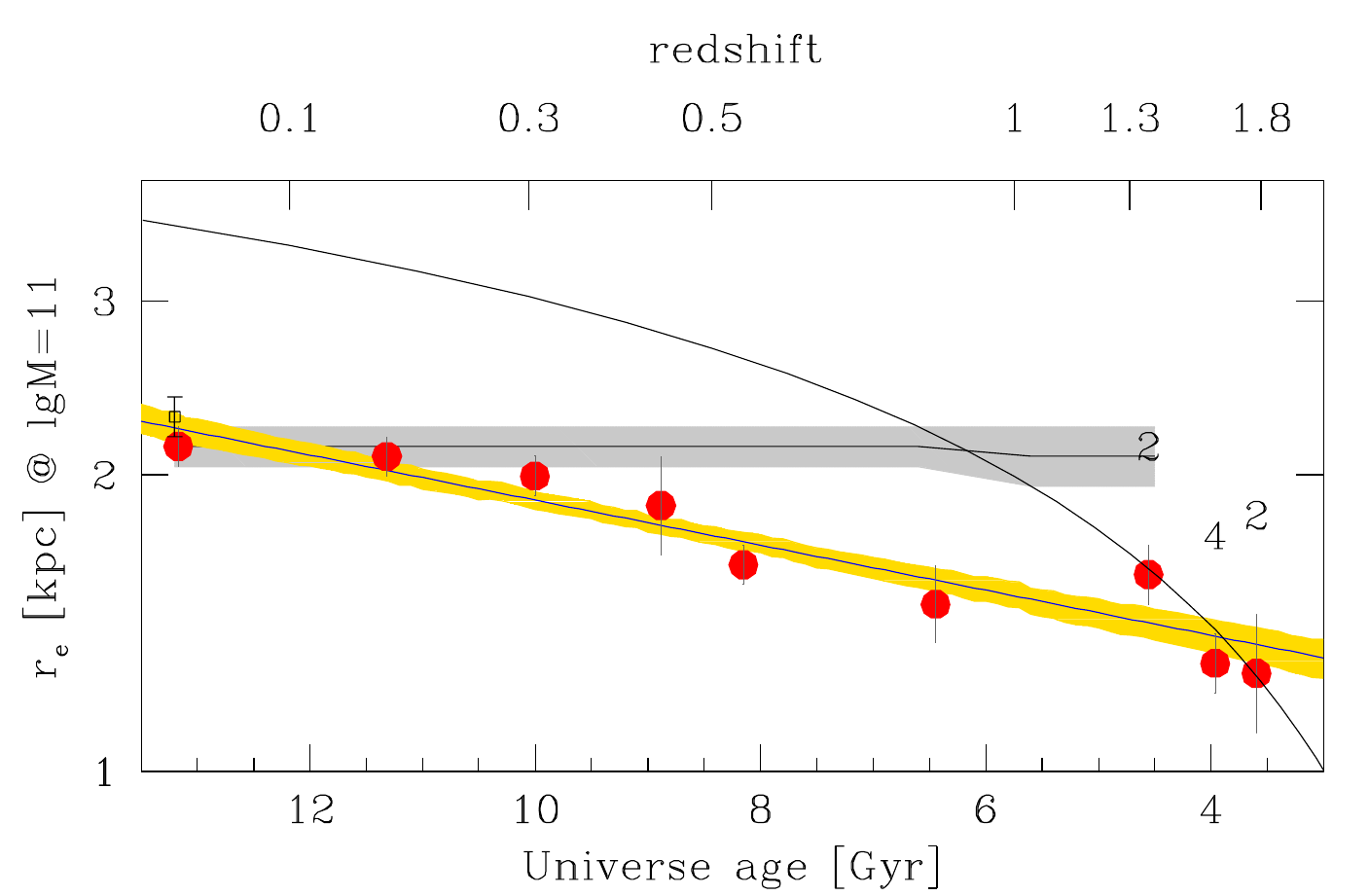}}
\caption[h]{Size at $\log M/M_\odot=11$ vs redshift. The number above the
points, when present, indicates the number of combined clusters. 
The solid line and shading show the fitted 
relation and its 68 \% uncertainty (posterior highest density interval).
The curve indicates the evolution measured in the field by Newman et al. (2012);
its absolute location is arbitrary and depends, among other things, on the adopted
initial mass function and the way 
the population under study is selected. The almost horizontal line indicates the 
effect of progenitor bias on Coma galaxies. Its shading indicates the
formal 68\% uncertainty. The open square point is our fit of the mass-size
relation using the data tabulated in Jorgensen et al. (1995, 1999) for early-type
Coma galaxies.
}
\end{figure}

\section{Discussion}

\begin{table*}
\caption{Mass-size fitting parameters: intercept $\gamma$, slope $\alpha$ and intrinsic
scatter $\sigma$ for the various samples}
\begin{tabular}{l l l l l}
\hline
Sample & \multicolumn{1}{c}{$\gamma$} & \multicolumn{1}{c}{$\alpha$} & \multicolumn{1}{c}{$\sigma$} & Comments \\  
\hline
$z=1.78$ &  $0.10\pm0.06$ & $0.61\pm0.55$ & $0.21\pm0.05$ &  $z=1.75-1.80$ \\
$z=1.60$ &  $0.11\pm0.03$ & $0.80\pm0.16$ & $0.15\pm0.02$ &  $z=1.48-1.71$ \\
$z=1.36$ &  $0.20\pm0.03$ & $0.57\pm0.12$ & $0.13\pm0.02$ &  $z=1.32-1.40$ \\ 
RXJ0152.7-1357 &  $0.17\pm0.04$ & $0.83\pm0.12$ & $0.18\pm0.02$ &  $z=0.84$ \\ 
MACSJ1149.5+2223 &  $0.21\pm0.02$ & $0.51\pm0.10$ & $0.10\pm0.01$ & $z=0.54$  \\
MACSJ1206.2-0847 &  $0.27\pm0.05$ & $0.43\pm0.14$ & $0.20\pm0.03$ & $z=0.44$  \\
Abell 2744 &  $0.30\pm0.02$ & $0.52\pm0.11$ & $0.10\pm0.02$ & $z=0.31$  \\
Abell 2218 & $0.32\pm0.02$ & $0.44\pm0.09$ & $0.11\pm0.02$ & $z=0.17$\\
Abell 1656 (Coma) &  $0.33\pm0.02$ & $0.60\pm0.06$ & $0.16\pm0.01$ & $z=0.02$  \\
\hline		   
\end{tabular}      				  
\hfill \break	      
\hfill \break	   
\end{table*}

\begin{figure}
\centerline{\includegraphics[width=6truecm]{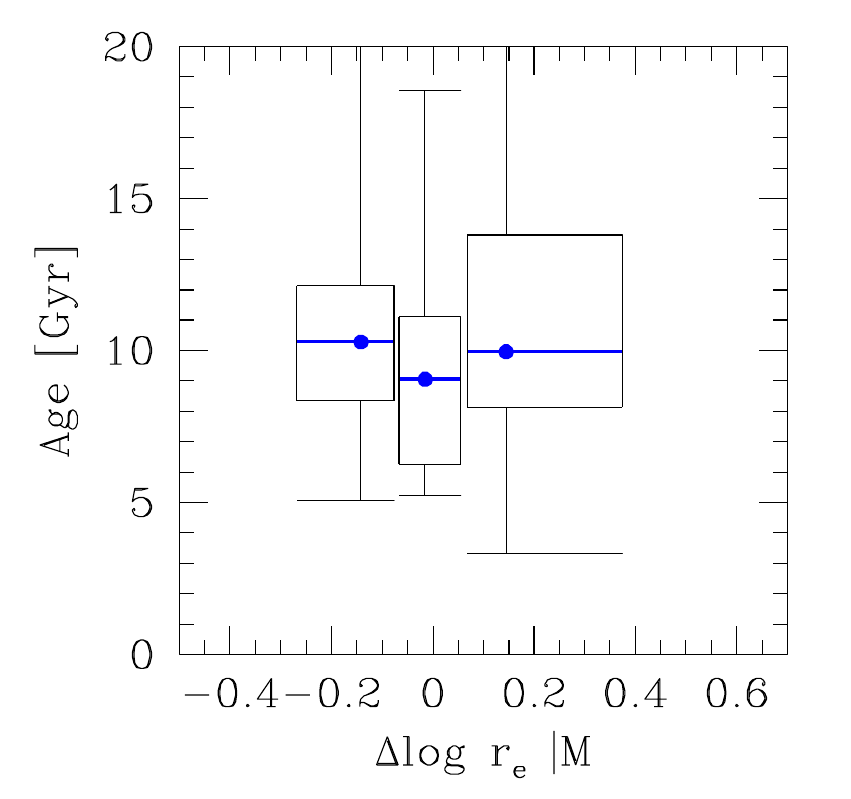}}
\caption[h]{Age distribution of Coma early-type galaxies 
smaller/average/larger for their mass. 
The plot is a standard box-whisker: the vertical box width
delimits the 1$^{st}$ and 3$^{rd}$ quartile, while the median (2$^{rd}$ quartile)
is indicated by the horizontal (blue) segment inside the box.
The horizontal box width the full x range of each bin,
while the error bars reach the minimum/maximum in the ordinate. 
Observed ages can be older than the Universe age because
of errors. 
}
\end{figure}

Before discussing our conclusions in the context of other works, we
need to remember two differences.
First, the half-light radius is the radius that enclosed half of the
galaxy luminosity and our analysis strictly adopts this
definition. Some other works adopt a different definition of 
galaxy size and, as discussed in Appendix B, these values should be
combined with or compared to our half-light radii with great caution.

Second, we classify galaxies following the
definitions of the morphological types. Some other works
sometimes adopt different
definitions for the morphological types, leading to samples 30\% to 50\% contaminated by
non-early-type galaxies, as detailed in Appendix C. Galaxy populations
selected with different criteria may well evolve differently.

Once considered these caveats,
our results agree with literature analysis of
galaxy cluster datasets, but 
constrain the size evolution more because of
a more extended and more
homogeneous sampling of the look-back time, a larger number of analyzed clusters, 
and more uniform morphological and half-light radii determinations.

\subsection{Secular or environmental processes?}

At first sight, the marked difference between the size evolution seen in
clusters and the larger increase in size seen in
field galaxies over the same 10 Gyr period (Fig.~14)
excludes that secular processes, such as
stellar winds and AGN, are primarily responsible of the galaxy radial growth because
these processes operate independently of environment.
However, this assumption strongly relies on the soundness of the derived
mass-size relations. Although various works
(e.g., Newman et al. 2012; van der Wel et al. 2014; Carollo et al. 2013) 
find consistent results for field samples, we note that first, these authors did not use
half-light radii, but scale-lengths and these can be compared with caution with our
radii because these authors
hold ellipticity and position angle fixed at all radii for galaxies known to have gradients
(see also Appendix B). On the other hand, by only comparing the size evolution, our comparison
is unaffected by a redshift-independent systematics. 
Second, samples are usually not morphologically classified 
as we did for clusters (i.e., by resemblance to standard), but by other criteria.
For example, van der Wel et al. (2014) do not morphologically
classify galaxies at all, whereas Carollo et al. (2013) morphological classify
galaxies using a support vector machine algorithm that in the case of the Delaye
et al. (2012) sample returned a sample very contaminated by late-type galaxies
(see Appendix C).
Different galaxy populations (all quiescents vs early-type quiescents)
may well evolve differently.
Third, morphological
misclassification is ignored, 
with the exception of Newman et al. (2014), who find a consistent
mass-size relation in the JKCS\,041 cluster and in the coeval field after accounting
for misclassification.  Fourth, it is puzzling that
galaxies have the same size in cluster and in
the field at $z=0$ (e.g., Pahre et al. 1998) and at $z=1.8$ (Newman et al. 2014),
but that the variation in size between these redshifts is different.
Therefore, while the marked difference between evolution in cluster
and field suggests excluding secular processes as being responsible
for the galaxy radial growth, differences in the way galaxies
are selected and sizes are measured preclude 
to draw a firm conclusion on which process shapes 
galaxy sizes.

\subsection{Collective and individual size evolution?}

At every redshift, we only consider galaxies that are quiescent and early-type
at the cluster redshift, ignoring those that will be as such at $z=0$.
We have therefore determined the size growth of the population that is
red and early-type at the redshift of observation. This determination is
valuable because it is well-defined and reproducible, and 
it describes the evolution of
a population selected at the redshift of observation.

If, however, galaxies become quiescent 
or morphologically early-type in high numbers in the last 10 Gyr,
the size evolution derived above may differ from the evolution
of the sample formed by the galaxies that will be quiescent 
and early-type at $z=0$. In this case, our study is
not comparing ancestors to descendents, an effect sometime called 
ancestor bias (Andreon \& Ettori 1999; van Dokkum \& Franx 2001). 
This is especially important for clusters, which have grown by a factor of
three in mass (and plausible in galaxy content) since $z\sim2$ (e.g., Fakhouri et al.
2010), which means that most of the $z=0$ galaxies are {\it not} in the high-redshift
sample (and especially so in the small field observed by HST),
a point already discussed in the context of the mass-size relation
in Andreon et al. (2014). In other
words, we may be observing an almost static mass-size relation as a result of
the balance of two galaxy populations, an older one evolving in one way
and a newly quenched population evolving in a way
to almost perfectly compensate the evolution of the older population.

Can we quantify the amount of newly quenched early-type galaxies, i.e., those
not selected by our criteria at the redshift of observation but that will be
at $z=0$? Newly quenched early-type galaxies may come from four possible reservoirs.
First, they can be blue at the cluster redshift. However, the fraction of blue galaxies is
small and depends little on
redshift except at masses lower than considered here (Raichoor \& Andreon
2012), especially in the inner region of the cluster studied
in our work because galaxies are mass-quenched at higher redshift
(than studied by these authors, i.e., $z<1.2$ with one exception).
Therefore the blue galaxy population seems to make no (or little) contribution to the
progenitor bias.
Second, the newly quenched early-type galaxies may be
inside our color selection but have a late-type morphology at the cluster redshift.
This possibility indeed occurs at both extremes of the redshift range;
there are well-known spiral galaxies on the Coma red sequence
(e.g., Andreon 1996; Terlevich et al. 2001) and spectroscopically
confirmed red-sequence but non-early-type galaxies in JKCS\,041 (Newman et al. 2014
and this work). Third, galaxies 
can have a mass lower than the mass threshold at the redshift of observation, 
but will have a
mass increase (e.g., because of star formation or minor mergers) 
that would make them to enter the sample at lower redshifts. This population is
minor in size because the galaxy mass function in cluster does not evolve
(e.g., De Propris et al. 1999, 2007; Andreon 2006, 2013), in particular those
of quiescent galaxies (Andreon 2013; Andreon et al. 2014).
Fourth, the newly quenched early-type galaxies may be outside the HST field
of view, typically 0.5 Mpc at high redshift.

The impact of the newly quenched population, irrespective of their location in
the color-magnitude plane or in space at the redshift
of observation, can be estimated from fossil evidence, i.e., by taking
an age-tagged mass-size relation in the local Universe and checking for a 
trend
between age and vertical offset from the mean mass-size relation. This is illustrated for
Coma early-type galaxies (morphologies from Andreon et al. 1996, 1997; 
spectroscopic ages from
Smith et al. 2012) with the whisker plot in Fig.~15 showing the
age distribution of galaxies smaller/average/larger for their mass (from left
to right in the figure). 
The median age is as large 
as the times considered in this work (10 Gyr) with little or no dependence
on the offset from the mean mass-size relation. More than
$50\%$ of larger galaxies for their mass are $>10$ Gyr old,
i.e., would be in our $z\sim1.5$ sample, and $>75\%$ of them are old enough
to be in our $z<1$ samples. The population of newly quenched galaxies
(i.e., young) is therefore too small numerically and not different enough in size
at a given mass to be able to alter the location of the
mean mass-size relation. Quantitatively, we fit the Coma mass-size excluding
from it galaxies with ages that would make them not selected at the various
cluster redshifts because
bluer than our threshold for inclusion (0.2 mag bluer than the red sequence). 
Figure 14 shows the result of this exercise: the mean galaxy size
at $\log M/M_\odot=11$ changes
by $0.01\pm0.03$ between today and $t=4.5$ Gyr, in agreement with
a similar computation on a reduced redshift range by
Jorgensen et al. (2014)\footnote{Some authors (e.g., Belli et al. 2015, Valentinuzzi et al. 
2010, Saglia et al. 2010) find a slighly larger progenitor bias, but for samples 
not morphologically selected, i.e., that include late-type galaxies.}.
The shading gives the negligible 68\% uncertainty 
ignoring errors on the age determinations of the individual
galaxies. Ages become less reliable 
with look-back time,and this is reason why we did not extrapolate past $t=4.5$ Gyr.
To summarize, judging from
what we see at $z=0$, the newly quenched population is numerically minor
and not different enough in size
at a given mass to be able to significatively alter the location of the
mean mass-size relation. Therefore, the size evolution
measured by selecting galaxies at the redshift of observation is indistinguishable
from the evolution that compares ancestors and descendents.

\section{Conclusions}

We carried out a photometric and structural analysis in the rest-frame $V$-band of 
a mass-selected ($\log M/M_\odot >10.7$) sample of red-sequence
galaxies in 14 galaxy clusters, 6 of which are at the key redshift where
most of the size evolution occurs in the field,
$z>1.45$. These are JKCS041, IDCS J1426.5+3508,
SpARCS104922.6+564032.5, SpARCSJ022426-032330, XDCPJ0044.0-2033, and
SPT-CLJ2040-4451. The other eight clusters uniformly sample the last 7 Gyr of the Universe age. 
To measure the size evolution of red-sequence early-type galaxies, 
we reduced and analyzed 
about 300 orbits of multicolor images taken by the 
Hubble Space Telescope. We uniformly
morphologically classified galaxies from $z=0.023$
to $z=1.803$, and we homogeneously derived sizes (half-light radii) for the 
entire sample. 
Our size derivation allows the presence of the variety of morphological
structures usually seen in early-type galaxies, such as bulges, bars, disks, isophote
twists, and ellipiticy gradients. For this reason, it is unbiased
by these structural components.
By using such a mass-selected 
sample, composed of 244 red-sequence
early-type galaxies, we find that the $\log$ of the galaxy size at a fixed stellar
mass has increased
with time with a rate of $0.023\pm0.002$ dex per Gyr over the last 10 Gyr,
in marked contrast with the threefold increase found in the literature
for galaxies in the general field over the same period. 
If we trust the field determination,
the marked difference between the size evolution seen in
clusters and in
field galaxies over the same 10 Gyr period 
excludes that secular processes, such as
stellar winds and AGN, are primarily responsible for the galaxy radial growth because
such processes operate independently of environment. However, 
differences in the way sizes are derived and samples are selected in the different
environments preclude any firm conclusion.
Using spectroscopic ages of Coma early-type
galaxies we also find that recently quenched early-type galaxies are a numerically
minor population not different enough in size to alter the mean size at a given
mass, which implies that the progenitor bias is minor, i.e., 
that the size evolution
measured by selecting galaxies at the redshift of observation is indistinguishable
from the one that compares
ancestors and descendents.

To put firmer constraints on the physical processes responsible for this size evolution,
this work could be extended in three interesting directions. First, by
duplicating a similar homogeneous analysis on a field sample 
to allow a fair comparison with our cluster sample. 
Second, by considering separately E and S0 galaxies, given the
debate on their possible differential evolution
(e.g.,  Dressler et al. 1997 vs Andreon 1998b, Lubin et al. 1998, Holden et al.
2009, Burstein et al. 2005). Third, by duplicating the progenitor bias
computation but reducing the extrapolation, i.e., using a sample
at $z\gg0$. 

\begin{acknowledgements}
SA thanks Richard Ellis, Drew Newman, and Tommaso Treu for the
paper's introduction, largely drawn from a proposal written together.
We thank Tommaso Treu/Chris Haines for the spectroscopic catalog of MACSJ1144/A2218, 
made available
in advance of publication. We thank Russell Smith for letting us use
the full Coma spectroscopic catalog in advance of publication.
We thank Saul Perlmutter for observing and making
immediately public the high-redshift clusters analyzed here. 
We also thank Sirio Belli, Chris Haines, Drew Newman, Roberto Saglia, Paolo Saracco, and
Veronica Strazzullo for comments on an earlier
version of this draft.
Finally, we thank
the HST archive (full acknowledgment is available at 
https://archive.stsci.edu/hst/acknowledgments.html ).
\end{acknowledgements}

{}

\appendix

\section{Comments on individual galaxies or clusters}

In the $z=1.803$ JKCS041 cluster, 
galaxies classified as non-early all are $>0.5$ mag bluer (except one case just $0.2$ mag
bluer) than red-sequence galaxies in $F606W-F814W$. This is a common theme for all
clusters, although galaxies are {\it morphologically} classified, their morphology 
is tightly correlated to color: non-early galaxies have blue colors.

One galaxy in the direction of the $z=1.75$ IDCS J1426.5+3508 cluster 
is redder than red-sequence galaxies in all filters (Fig.~9), and increasingly so
going toward bluer filters (e.g., $0.7$ mag in F606W-F814W), making it a very likely cluster
non-member. For this reason it is excluded from the sample. Furthermore, it lies at a crowded
location, making the isophotal analysis of this object almost impossible. 

The $z=1.71$ SpARCS104922.6+564032.5 cluster
is a poor cluster, only providing
three red-sequence early-type galaxies.

The $z=1.63$ SpARCSJ022426-032330 cluster 
is also a poor cluster.
Its brightest galaxy (ID=1207)  has a blue F105W-F140W
color (Fig.~9), but a normal F814W-F105W color. A 13 ks XMM exposure (from the archive)
shows no X-ray source coincident with this galaxy. Our isophotal analysis shows this
galaxy to be an outlier in the mass-size relation (too small for its mass, Fig.~12)
and is neglected in the mass-size fit (see Fig.~13). 
As for IDCS J1426.5+3508, all red-sequence non-early-type galaxies
are bluer in F814W-F105W than red-sequence early-type galaxies, with no exceptions. 
The non-analyzable galaxy (open point in Figure~9) is
fully embedded in the brightest cluster galaxy.

Morphology and color match each other in the $z=1.58$ XDCPJ0044.0-2033 
cluster with only one exception.

A large nearby irregular galaxy partially hides the $z=1.48$ SPT-CL2040-4451 
cluster (see Fig.~5). 
We masked this large galaxy to improve background subtraction of the remaining part
of the image.

Out of 22 early-type galaxies on
the RXJ0152.7-1357 red sequence, 14 have a spectroscopic redshift 
and only one is outside the cluster, confirming that almost all 
of the analyzed galaxies are members.

Out of 24 photo-z selected early-type galaxies on
the MACSJ1149.5+2223 red sequence, 2 have unfeasible isophotal analysis. 
Of the remaining 22 galaxies,
17 have a spectroscopic redshift and are 
all spectroscopically confirmed members. The size of 
one galaxy (a spectroscopic member) is too small for its mass, and is
neglected in the mass-size fit (see Fig.~13).

As for IDCS J1426.5+3508, all the few F814W-F606W red-sequence non-early-type galaxies of Abell 2744
are bluer in F435W-F606W than red-sequence early-type galaxies, with no exceptions.  Furthermore,
they are all spectroscopically confirmed non-members. Of the 38 remaining galaxies, 31 have
a spectroscopic redshift, and all but one are spectroscopically confirmed members.
The Abell 2744 cluster is extraordinary
rich and massive, which makes its core crowded. We therefore exclude from our analysis 14 
early-type galaxies with largely overlapping isophotes in the very
center of the cluster.  

All but one galaxy in Abell 2218 has a spectroscopic redshift, and all turn out to be
cluster members. A few galaxies with very overlapping isophotes at the cluster center
(plus a galaxy pair) have been ignored in our isophotal analysis.

\section{Half-light radius vs scale-length}

\begin{figure}
\centerline{\includegraphics[width=5truecm]{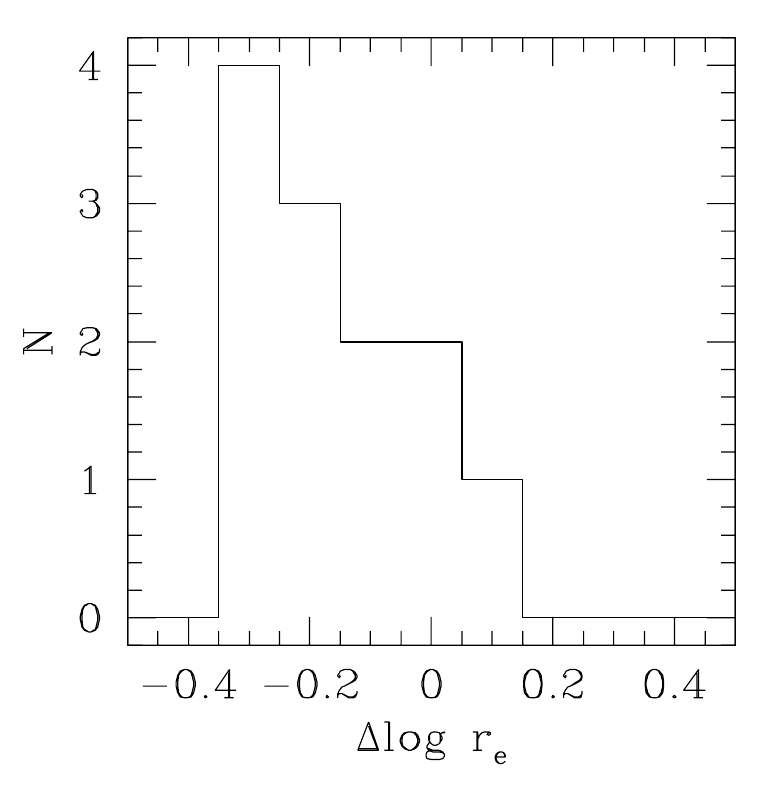}}
\caption[h]{Size difference (ours minus literature) for 12 early-type galaxies
in common between our work and Blakeslee et al. (2006).
}
\end{figure}

The half-light radius is the radius of the isophote that encloses half of the
galaxy luminosity and our analysis strictly adopts this
definition. As mentioned in Sec.~3.1, there is a large body of literature
that uses half-light radii and that has compared different code implementations
to derive it. Many recent works  
assume that galaxies are single-component objects, compute the
azimuthally averaged radial profile and fit it with a
radial profile to derive the
scale--length parameter, often called effective 
radius (or major semi-axis) by some GALFIT (Peng et al. 2002) users. It is
one of the parameters describing the shape of the azimuthally averaged
radial profile and 
only encloses half the galaxy light for radial profile perfectly described
by the assumed (Sersic) radial profile at all radii 
(Djorgovski \& Davies 1987) and for single-component galaxies,
i.e., for galaxies without any radial change in position angle or ellipticity, as also remembered
by Peng et al. (2002, 2010). However, almost no galaxy fits this description,
as discussed in the introduction.
Because the fitted model is
an oversimplified description of nature complexity,
the scale-length so derived
is expected to differ
from the half-light radius (or semimajor axis). 
Instead, our half-light derivation allows the full complexity of
the morphological structure seen in early-type galaxies, including 
ellipticity or position angle gradients. We note that although the new version of
GALFIT (Peng et al. 2010) allows a multi-component fit, it has never
been used in our context to our best knowledge, except Lang
et al. (2014). This work shows that at least two-thirds of the galaxies
in CANDELS/3D-HST requires a multi-component fit.

Given that we do not use scale-lengths in our work, a detailed study of the relative
differences between half-light radius and scale-length is beyond the scope of the work. 
Nevertheless, 
RXJ0152.7-1357 early-type galaxies in our work, in Delaye et al. (2014), Blakeslee et al. (2006), 
and Chiboucas et al. (2009), having up to three scale-length estimates based
on single-component GALFIT fits and one half-light radius,
can be used to have an idea of the possible differences among these two quantities.
Figure~B.1 shows for 12 galaxies in common
with Blakeslee et al. (2006) that
scale lengths derived from azimuthally averaged profiles
show a scatter with half-light radius
and that are 
larger, on average, than half-light radii. 
The sign of the difference is expected because at large radii an ellipse with 
incorrect position angle
or ellipticity intercepts more flux than an ellipse fitting the isophote.
This leads to
a shallower profile and therefore to larger scale-lengths (see also Lang et al. 2014).
The amplitude of the systematic depends on many factors related
both to the precise GALFIT setting (the three GALFIT runs return different values
for some galaxies, see also Chiboucas et al. 2009) and to galaxies themselves, such
as the photometric importance of the disk, and/or amplitude of the isophotal twist.
Therefore, the systematic is hard to predict, making difficult
to combine scale-length determinations to our half-light radii.
The smaller samples in
common with Delaye et al. (2014), and Chiboucas et al. (2009) (8 and 5 galaxies) confirm
similar offsets. Carollo et al. (2013) derived sizes using
a different software that however  
holds ellipticity and position angle
fixed at all radii. Therefore, this work also likely shares the 
shortcomings of a too simple modelling.

To summarize, 
scale-lengths of a
galaxy model that is too simple to fit the known features 
(e.g., gradients in ellipticity or
position angle) of early-type
galaxies at both high and low redshift  are not half-light radii.

\section{Morphological classification}

\begin{figure}
\centerline{%
\includegraphics[width=3truecm]{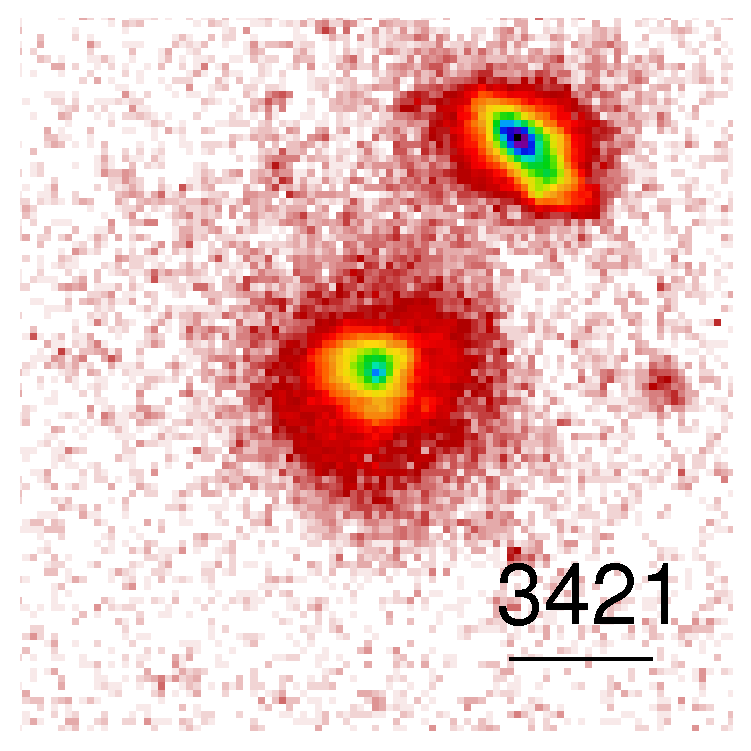}%
\includegraphics[width=3truecm]{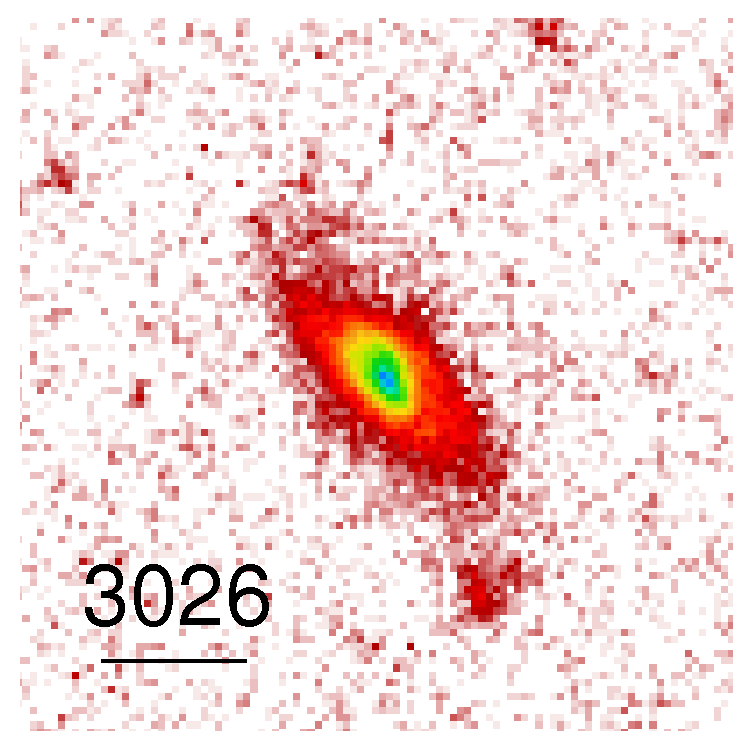}%
\includegraphics[width=3truecm]{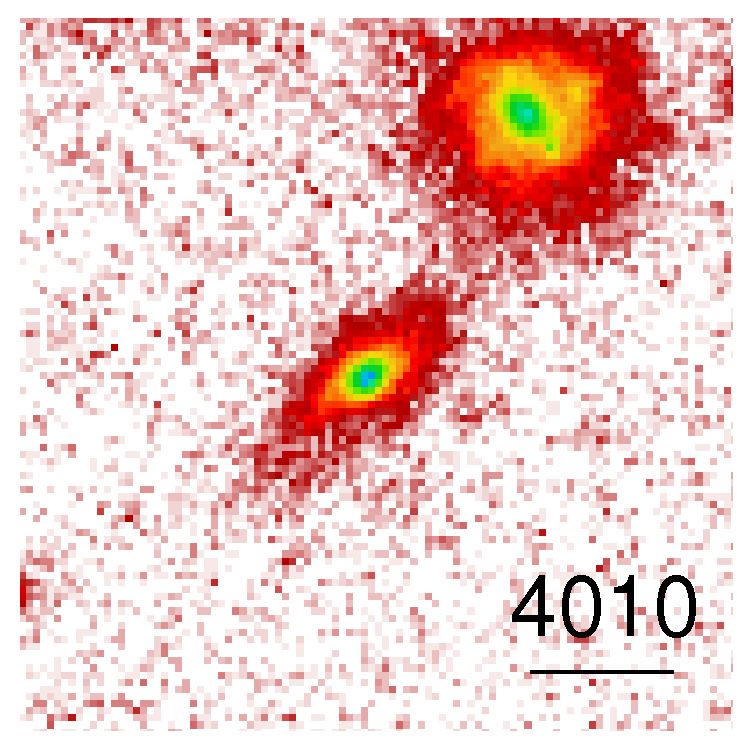}%
}
\centerline{%
\includegraphics[width=3truecm]{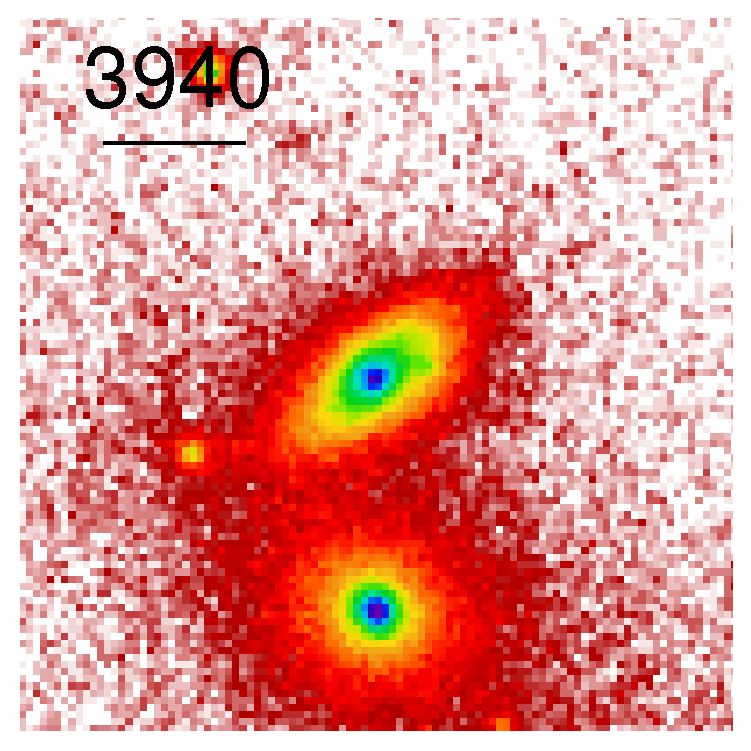}%
\includegraphics[width=3truecm]{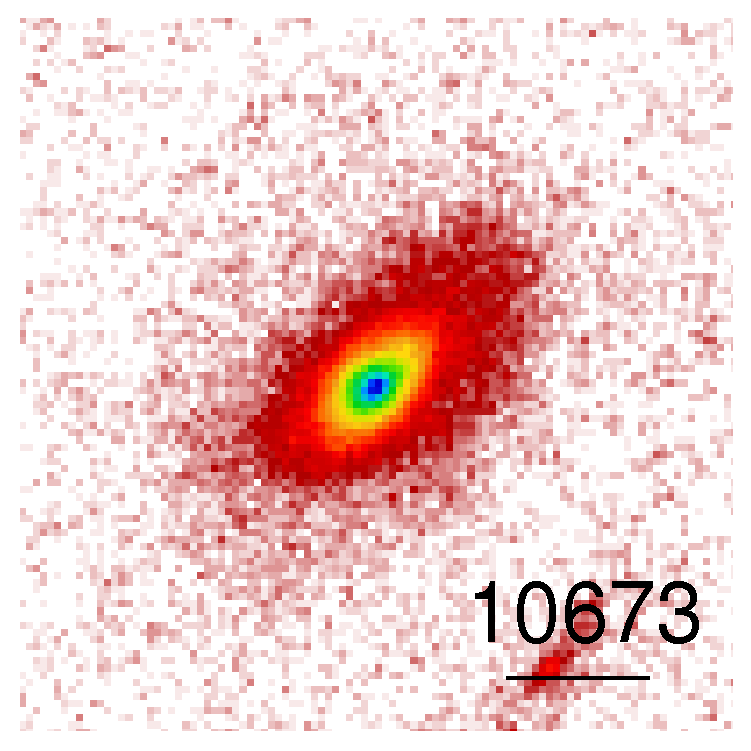}%
\includegraphics[width=3truecm]{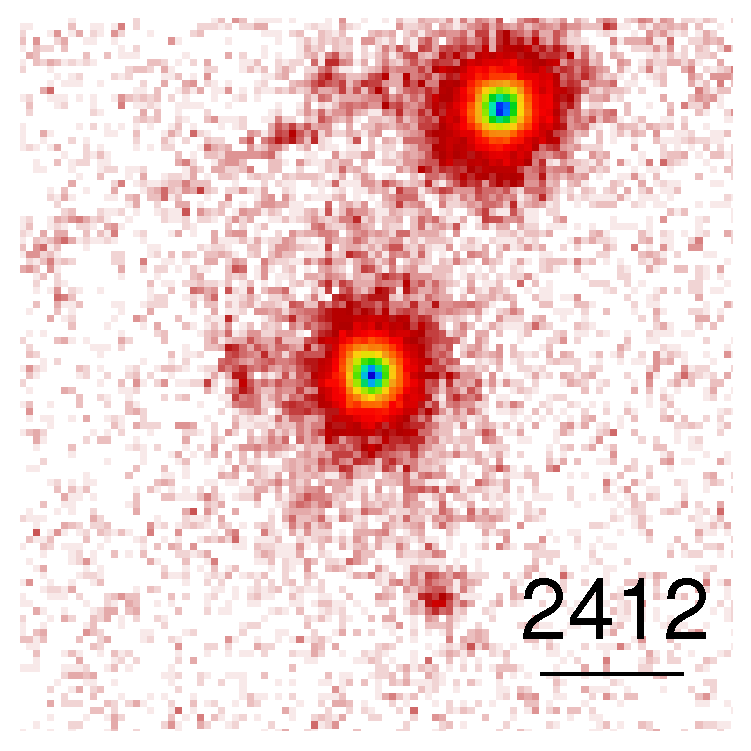}%
}
\caption[h]{Remarkable cases of late-type galaxies previously
classified as early-type. These galaxies have manifestly
irregular or S-shaped isophotes. Top and bottom panels
show examples in Delaye et al. (2014) and Blakeslee et al. (2006),
respectively. The tick is 1 arcsec. The numbers are tthe IDs in the respective papers.
}
\end{figure}

We classify 
galaxies following the
definitions of the morphological types. As detailed in Sec.~3, our
classification returns morphologies coincident with the one derived by
morphologists. Some other works
sometimes adopt different morphological classes, and
it may happen that galaxy populations selected in different ways
also evolve in different ways.

RXJ0152.7-1357
galaxies, studied by us, Delaye et al. (2014), and Blakeslee et al. (2006),
offer a sample useful to understand classification systematics.
We first
emphasize that Delaye et al. (2014) use a support vector machine algorithm
to classify the galaxies, whereas Blakeslee et al. (2006) use morphologies
from Postman et al. (2005), which are based on eye inspection and 
qualitative resemblance to morphological standard. 
We morphologically re-classify the galaxies in both the Delaye et al. (2014) 
and Blakeslee et al. (2006) samples, finding that
the Delaye et al. (2014) sample is 50\% contaminated by late-type galaxies, 
while the Blakeslee et al. (2006) sample is about 30\%
contaminated. Some remarkable examples are shown in Fig.~C.1.
Since the compositions of the three `early-type' samples are quite
different, differences in the mass--size relation and evolution
may arise.  
A similar morphological comparison exercise for a $z=0.4$
cluster is shown in Andreon (1998b), finding again that
non-early-type galaxy features are sometime cumbersome to detect by
visual inspection, but hard to escape by measuring 
isophote shapes (and also visible by eye inspection once a
proper display is found, as in Fig.~C.1).

The morphological complexity of the contaminating
population worsen the performances in recovering half-light sizes 
of those programs that assume galaxies to be single components because
late-type galaxies markedly differ from the single component assumed by the fitted model. Finally,
the often large contamination is usually counted by authors 
as an additional signal rather than as contamination, spuriously overestimating
the quality of their measurement.

\section{Comparison with previous cluster works}

Compared to literature analysis, our work displays a more extended and more
homogeneous sampling of the look-back time, a larger number of analyzed clusters, 
and morphological classification and half-light radii determination that are
more uniform and compliant to the definitions. Once considered the morphological and size
caveats discussed in the previous sections, our results are consistent with the
literature results, but constraining more tightly the size evolution, or differ
for understood reasons detailed below.

For example, Saracco et al. (2014) find consistent mass-size relation comparing
Coma and a $z=1.26$ cluster, similarly to the more
constraining work of Andreon et al. (2014) that has an enlarged redshift
baseline. Both works lack sensitivity, being based on a handful of
galaxies at high redshift, and indeed the more stringent upper limit derived in 
Andreon et al. (2015) is fully consistent with the change we measure
in the present work.

Delaye et al. (2014), as do the previous works, sparsely sample the 
redshift range, with a large gap between the $z\sim0$ point, taken from Huertas-Company
et al. (2013) and the next point, RXJ0152.7-1357 at $z=0.83$,
also present in our cluster sample. They
find a large evolution between these redshifts, but
their claim largely relies on their
choice of comparing their high-redshift very-massive clusters to a
zero-redshift sample of galaxies
mostly in low-mass clusters observed in shallow
and low-resolution images (the Sloan Digital Sky Survey). We verified
that the evidence of an evolution largely vanishes when adopting   
effective radii determined on deeper
and better seeing images of massive nearby clusters,
for example those
of Coma cluster galaxies published in 
Saglia et al. (1993a), Jorgensen et al. (1995, 1999),
and Andreon et al. (1996, 1997).
Papovich et al. (2012) compares instead a handful of galaxies in
an high-redshift group of mass comparable to our own Galaxy
(Tran et al. 2015) to a couple of massive clusters at intermediate redshift.

De Propris et al. (2015b) extend 
their previous work (De Propris et al. 2015a) on $z\sim1.25$ clusters
to the CLASH cluster sample 
however without separating galaxies in morphological classes (they
use the Sersic index). Because their sensitivity is at least
two times worse than our measured signal, 
they found no evolution at $z<0.6$, consistent
with the evolution we find. They found an evolution at
higher redshift, but we re-emphasize the use of a different classification
scheme (Sersic index vs morphology) and the change in the
morphological composition, also emphasized by the authors.
Our clusters at high
redshift have $\gtrsim 5$ times longer exposure times and more ditherings
than those in  
De Propris et al. (2015a), resulting in deeper images that are better sampled. 
In particular, dithering is paramount because many of the
De Propris et al. (2015a) galaxies have effective radii equal to one natural
WFC3 pixel. Indeed, their $z\sim1.25$ clusters are not in our sample because
they have images of insufficient quality for our standards.

La Barbera et al. (2003) also find no evolution in the Kormendy relation
once luminosity evolution is accounted for because of lack of sensitivity
of the used data.
In fact, three out of four of their clusters are at low redshift 
(two in common with the present work), while
the galaxies in the highest redshift cluster, at $z=0.64$, have effective sizes 
derived from ground-based imaging, which are challenging to derive with such 
low-resolution data. Finally,
their sample is not morphologically selected because this measurement
is inaccessible from their ground images.

\end{document}